\RequirePackage{ifpdf}
\ifpdf 
\documentclass[pdftex]{sigma}
\else
\documentclass{sigma}
\fi

\newcommand{\mR}{\mathbb{R}}
\newcommand{\mC}{\mathbb{C}}
\newcommand{\mN}{\mathbb{N}}

\newcommand{\mZ}{\mathbb{Z}}
\newcommand{\mS}{\mathbb{S}}

\newcommand{\mB}{\mathbb{B}}
\newcommand{\mV}{\mathbb{V}}
\newcommand{\mF}{\mathbb{F}}

\newcommand{\cH}{\mathcal{H}}

\newcommand{\cF}{\mathcal{F}}
\newcommand{\cP}{\mathcal{P}}

\newcommand{\cS}{\mathcal{S}}

\newcommand{\cL}{\mathcal{L}}

\newcommand{\cO}{\mathcal{O}}
\newcommand{\cV}{\mathcal{V}}
\newcommand{\cU}{\mathcal{U}}

\newcommand{\ux}{\underline{x}}

\newcommand{\uy}{\underline{y}}

\numberwithin{equation}{section}

\begin{document}

\allowdisplaybreaks

\renewcommand{\thefootnote}{$\star$}

\renewcommand{\PaperNumber}{047}

\FirstPageHeading

\ShortArticleName{The Fourier Transform on Quantum Euclidean Space}

\ArticleName{The Fourier Transform on Quantum Euclidean Space\footnote{This paper is a
contribution to the Special Issue ``Relationship of Orthogonal Polynomials and Special Functions with Quantum Groups and Integrable Systems''. The
full collection is available at
\href{http://www.emis.de/journals/SIGMA/OPSF.html}{http://www.emis.de/journals/SIGMA/OPSF.html}}}

\Author{Kevin COULEMBIER}

\AuthorNameForHeading{K. Coulembier}

\Address{Gent University, Galglaan~2, 9000 Gent, Belgium}
\Email{\href{mailto:Coulembier@cage.ugent.be}{Coulembier@cage.ugent.be}}
\URLaddress{\url{http://cage.ugent.be/~coulembier/}}

\ArticleDates{Received November 19, 2010, in f\/inal form April 21, 2011;  Published online May 11, 2011}

\Abstract{We study Fourier theory on quantum Euclidean space. A modif\/ied version of the general def\/inition of the Fourier transform on a quantum space is used and its inverse is constructed. The Fourier transforms can be def\/ined by their Bochner's relations and a new type of $q$-Hankel transforms using the f\/irst and second $q$-Bessel functions. The behavior of the Fourier transforms with respect to partial derivatives and multiplication with variables is studied. The Fourier transform acts between the two representation spaces for the harmonic oscillator on quantum Euclidean space. By using this property it is possible to def\/ine a~Fourier transform on the entire Hilbert space of the harmonic oscillator, which is its own inverse and satisf\/ies the Parseval theorem.}


\Classification{17B37; 81R60; 33D50}

\renewcommand{\thefootnote}{\arabic{footnote}}
\setcounter{footnote}{0}

\section{Introduction}
There has been a lot of interest in formulating physics on noncommutative space-times, see e.g.~\cite{MR1243712, MR1097174, MR1239953, MR1328733, MR1303081, MR1317458}. In particular, since non-commutativity implies a quantized space-time, quantum f\/ield theories on such spaces should be well-behaved in the ultraviolet-limit, see e.g.~\cite{MR1243712}. The inf\/inities of the commutative, continuous theories could appear as poles in the $q$-plane with $q$ a deformation parameter. In such theories quantum groups replace Lie groups in the description of the symmetries. An important concept in this theory is integration and Fourier theory on quantum spaces, see e.g.~\cite{MR1722348, MR1303081, MR1317458, MR1408120, MR2035025, MR2099760}. The Fourier kernel is def\/ined in~\cite[Def\/inition~4.1]{MR1235979}. In this paper we study the Fourier theory on quantum Euclidean space, which has symmetry group $O_q(m)$. The deformation parameter $q$ is always assumed to satisfy $0<q<1$. The Fourier transform is studied from the point of view of harmonic analysis on quantum Euclidean space, see e.g.~\cite{MR1097174, MR1328733, MR2006509, MR1182031, MR1361741}. This is captured in the Howe dual pair $(O_q(m),\cU_q(\mathfrak{sl}_2))$. The quantum algebra $\cU_q(\mathfrak{sl}_2)$ is generated by the $O_q(m)$-invariant norm squared and Laplace operator on quantum Euclidean space. The Fourier transform was def\/ined in an abstract Hopf-algebraic setting in~\cite{MR1303081}. In this article, the Fourier transform on quantum Euclidean space is studied analytically. This leads to explicit formulae for the behavior of the Fourier transform with respect to partial derivatives. The def\/inition of the Fourier transform is also extended from spaces of polynomials weighted with Gaussians to an appropriate Hilbert space, which was a problem left open in~\cite{MR1303081}.

A general theory of Gaussian-induced integration on quantum spaces was developed in~\cite{MR1303081}. We use this procedure on quantum Euclidean space for the two types of calculus def\/ined in~\cite{MR1138053}. One of the two integrations we obtain corresponds to the result in~\cite{MR1239953, MR1408120}. Both types of integration can be written as a combination of integration over the quantum sphere, see~\cite{MR1408120}, and radial Jackson integration, see e.g.~\cite[Section~1.11]{MR1052153}. Each one of the integrations satisf\/ies Stokes' theorem for both types of calculus. It turns out that Fourier theory is def\/ined more naturally using the Fourier kernel for one calculus combined with the Gaussian-induced integration for the other calculus. This implies we use a generalized Gaussian-induced integration compared to~\cite{MR1303081}. This was also done implicitly for the analytical approach to Fourier theory on the braided line in~\cite{MR1448685}. We calculate the quantum sphere integral of spherical harmonics weighted with the Fourier kernel, which yields a $q$-deformed Bessel function. This function is known as the f\/irst $q$-Bessel function, see~\cite{MR0649849}. As a side result we obtain a Funk--Hecke theorem on quantum Euclidean space. This allows us to construct the reproducing kernel for the spherical harmonics. The reproducing kernel can be expressed as a $q$-Gegenbauer polynomial in terms of the generalized powers of the inner product constructed in~\cite{MR1317458}.

Because of the appearance of $q$-Bessel functions, the combination of radial integration with the spherical integration and the exponential leads to new $q$-deformed Hankel transforms. In~\cite{MR1069750} the $q$-Hankel transforms corresponding to the so-called third $q$-Bessel functions were def\/ined and studied. In the current paper the $q$-Hankel transforms for the f\/irst and second $q$-Bessel functions are introduced. It is proven that they are each other's inverse by applying the theory of the $q$-Laguerre polynomials, see~\cite{MR0618759}. Then the inverse of the Fourier transform on quantum Euclidean space is def\/ined by its Bochner's relations in terms of the second $q$-Hankel transform. The fact that the Fourier transforms can be expressed in terms of Bochner's relations is an immediate consequence of their $O_q(m)$-invariance. It is proven that the transforms behave canonically with respect to partial derivatives and multiplication with variables.

Furthermore, we extend the domain of the Fourier transforms from spaces corresponding to polynomials weighted with a Gaussian to the Hilbert space structure of \cite{MR1239953}. This Hilbert space has two representations in function spaces; the Fourier transform and its inverse act between these spaces. The f\/irst and second $q$-Bessel function can be connected by a substitution $q\leftrightarrow q^{-1}$. The f\/irst $q$-Bessel function has a f\/inite domain of analyticity contrary to the second one. This implies that the inverse Fourier transform is better suited to generalize to a Hilbert space. By composing this Fourier transform with the projection operators corresponding to the two dual representations of the Hilbert space, we obtain a Fourier transform which can be def\/ined on the entire Hilbert space. This transform is its own inverse and satisf\/ies a Parseval theorem.

In \cite{MR2595267} the theory of the $q$-Dirac and $q$-Laplace operator on undeformed Euclidean space was developed. The $q$-Laplace operator is $O(m)$-invariant and generates $\cU_q(\mathfrak{sl}_2)$ together with the classical norm squared. This implies that $q$-harmonic analysis on Euclidean space corresponds to the Howe dual pair $(O(m),\cU_q(\mathfrak{sl}_2))$, i.e.\ there is no spherical deformation and the radial deformation corresponds to that of quantum Euclidean space. Therefore, the $q$-Hankel transforms in the current paper can also be used to construct an $O(m)$-invariant $q$-Fourier transform on Euclidean space, connected to the $q$-Dirac operator.

\looseness=-1
The paper is organized as follows. First an introduction to $q$-calculus, quantum Euclidean space and Fourier theory on quantum spaces is given. Then two $q$-Hankel transforms are def\/ined. By studying their behavior with respect to the $q$-Laguerre polynomials it is proven that the two transforms act as each other's inverse. Then the integration on quantum Euclidean space is studied. The Fourier transform of a spherical harmonic weighted with a radial function can be expressed as the f\/irst $q$-Hankel transform of the radial function. The inverse Fourier transform is therefore def\/ined by its Bochner's relations. Next, the behavior of the Fourier transforms with respect to derivatives and multiplication with variables is studied. The previous results allow a construction of a Funk--Hecke theorem and reproducing kernels for the spherical harmonics on quantum Euclidean space. Then the Fourier transforms are connected with the harmonic oscillator which makes it possible to extend the Fourier transform to the Hilbert space def\/ined for this harmonic oscillator. Finally the $q$-Fourier transform on undeformed Euclidean space is considered.

\section{Preliminaries}
\subsection[$q$-calculus]{$\boldsymbol{q}$-calculus}

We give a short introduction to $q$-derivatives, $q$-integration and $q$-special functions, see \cite{MR1052153, JACKSON, Koekoek, MR1448685}. The report \cite{Koekoek} that will be referred to often is also included in the book \cite{MR2656096}. For $u$ a~number, and $q$ the deformation parameter, $0<q<1$, we def\/ine the $q$-deformation of $u$ as
\begin{gather*}
[u]_q=\frac{q^u-1}{q-1}.
\end{gather*}
It is clear that $\lim\limits_{q\to 1}[u]_q=u$. We also def\/ine
\begin{gather*}
(u;q)_k=(1-u)(1-qu)\cdots\big(1-q^{k-1}u\big)\qquad \mbox{and}\qquad (u;q)_\infty=\prod_{k=0}^\infty \big(1-uq^k\big).
\end{gather*}

The $q$-derivative of a function $f(t)$ is def\/ined by
\begin{gather}
\label{qder}
\partial^q_t (f(t)) = \frac{f(qt)-f(t)}{(q-1)t}.
\end{gather}
This operator satisf\/ies the generalized Leibniz rule
\begin{gather}
\label{Leibniz}
\partial_t^q(f_1(t)f_2(t)) = \partial_t^q(f_1(t))f_2(t)+f_1(qt)\partial_t^q(f_2(t)).
\end{gather}

The $q$-integration on an interval $[0,a]$ with $a\in \mR$ is given by
\begin{gather*}
\int_{0}^af(t)\,d_qt = (1-q)a\sum_{k=0}^\infty f\big(q^ka\big)q^k.
\end{gather*}
The inf\/inite $q$-integral can be def\/ined in several ways, determined by a parameter $\gamma\in\mR\backslash \{0\}$,
\begin{gather}
\label{gammaint}
\int_0^{\gamma\cdot\infty}f(t)d_qt = (1-q)\gamma\sum_{k=-\infty}^\infty f\big(q^k\gamma\big)q^k=\lim_{l\to+\infty}\int_0^{q^{-l}\gamma}d_qt f(t).
\end{gather}
The positive $(\gamma\in\mR^+)$ or negative $(\gamma\in\mR^-)$ inf\/inite integral is a function of $\gamma$, however from the def\/inition it is clear that $\int_0^{\gamma\cdot\infty}=\int_0^{q\gamma\cdot \infty}$. So $\partial_\gamma^q\int_0^{\gamma\cdot\infty}d_qt=0$ holds which means the integral is a $q$-constant. The integral is the inverse of dif\/ferentiation,
\begin{gather}
\label{partint1}
\int_0^{a}\left[\partial_t^qf(t)\right]d_qt = f(a)-f(0).
\end{gather}

The $q$-factorial of an integer $k$ is given by $[k]_q!=[k]_q[k-1]_q\cdots [1]_q$ and satisf\/ies $[k]_q!=(q;q)_k/(1-q)^k$. This can be generalized to the $q$-Gamma function $\Gamma_q(t)$ for $t>0$ satisfying $\Gamma_q(t+1)=[t]_q\Gamma_q(t)$, see e.g.~\cite[formula (1.10.1)]{MR1052153}. The $q$-exponentials are def\/ined as
\begin{gather*}
e_q(t)=\sum_{j=0}^\infty\frac{t^j}{[j]_q!} \qquad \mbox{and} \qquad E_q(t)=e_{q^{-1}}(t)=\sum_{j=0}^\infty q^{\frac{1}{2}j(j-1)}\frac{t^j}{[j]_q!}.
\end{gather*}
Note that a dif\/ferent notation for the exponentials is used compared to \cite{MR2595267}. The relation $e_q(t)E_q(-t)=1$ holds and the derivatives are given by
\begin{gather}
\label{aflexp}
\partial_t^qe_q(t)=e_q(t) \qquad \mbox{and}\qquad \partial_t^qE_q(t)=E_q(qt).
\end{gather}
For $q<1$ the series $E_q(t)$ converges absolutely and uniformly everywhere and $e_q(t)$ in the area $|t|<\frac{1}{1-q}$. The function $e_q(t)$ can be analytically continued to $\mC\backslash \{\frac{q^{-k}}{1-q}\}$ as $1/(E_q(-t))$. The zeroes of the $q$-exponential $E_q$ are
\begin{gather}
\label{zeroes}
E_{q}\left(-\frac{q^{-k}}{1-q}\right) = 0\qquad \mbox{for} \ \ k\in\mN.
\end{gather}
This follows from the inf\/inite product representation $E_q(\frac{t}{1-q})=(-t;q)_\infty$, see \cite[formula (1.3.16)]{MR1052153}. This implies that the relation
\begin{gather}
\label{intinfeind}
\int_0^{\frac{1}{\sqrt{1-q}}\cdot\infty}d_qt f(t) E_{q^2}\left(\frac{-t^2}{1+q}\right)=\int_0^{\frac{1}{\sqrt{1-q}}}d_qt f(t) E_{q^2}\left(\frac{-t^2}{1+q}\right)
\end{gather}
holds.

The $q$-Hermite polynomials are given by
\begin{gather}
\label{Hermite}
H_k^q(t) = \sum_{j=0}^{\lfloor k/2\rfloor}(-1)^j\frac{[k]_q!}{[k-2j]_q![j]_{q^2}!}q^{j(j+1)}((q+1)t)^{k-2j}
\end{gather}
and related to the discrete $q$-Hermite~I polynomials $h_k(x;q)$ in \cite[Section~3.28]{Koekoek}  by
\begin{gather*}
H_k^q(t) = q^k\left(\frac{1+q}{1-q}\right)^{k/2}h_k\Big(q^{-1}\sqrt{1-q^2}t;q\Big).
\end{gather*}

We introduce the $q$-Laguerre polynomials for $\alpha>-1$ in the normalization of \cite[p.~24]{MR2595267},
\begin{gather}
\label{qLaginv}
\cL^{(\alpha)}_j\big(u|q^{-2}\big) = q^{-j(j+1+2\alpha)}\sum_{i=0}^jq^{2i(i+\alpha)}\frac{(-u)^i}{[j-i]_{q^2}![i]_{q^2}!}\frac{(q^{2i+2\alpha+2};q^2)_{(j-i)}}{(1-q^2)^{j-i}}.
\end{gather}
They can also be def\/ined using the $q$-Gamma function since $\frac{(q^{2i+2\alpha+2};q^2)_{(j-i)}}{(1-q^2)^{j-i}}=\frac{\Gamma_{q^2}(j+\alpha+1)}{\Gamma_{q^2}(i+\alpha+1)}$ holds. They are connected with the $q$-Laguerre polynomials $L_j^{(\alpha)}(u;q)$ from \cite[Section~3.21]{Koekoek}  by
\begin{gather*}
\cL_j^{(\alpha)}\big(u|q^{-2}\big) = q^{-j(j+1+2\alpha)}L_j^{(\alpha)}\big(\big(1-q^2\big)u;q^2\big)
\end{gather*}
and to the $q$-Laguerre polynomials in \cite{MR0618759} by the same formula with a substitution  $(1-q^2)u\to u$ in the right hand side of the formula.

The substitution $q\to q^{-1}$ yields,
\begin{gather}
\label{qLag}
\cL^{(\alpha)}_j\big(u|q^2\big) = \sum_{i=0}^jq^{(j-i)(j-i+1)}\frac{(-u)^i}{[j-i]_{q^2}![i]_{q^2}!}\frac{(q^{2i+2\alpha+2};q^2)_{(j-i)}}{(1-q^2)^{j-i}}.
\end{gather}
These polynomials are related to the little $q$-Laguerre polynomials (the Wall polynomials) $p_j(u;a|q)$, see \cite[Section~3.20]{Koekoek}, by
\begin{gather*}
\cL^{(\alpha)}_j\big(u|q^2\big) = q^{j(j+1)}\frac{(q^{2\alpha+2};q^2)_j}{(1-q^2)^j[j]_{q^2}!}p_j\left(\big(q^{-2}-1\big)u;q^{2\alpha}|q^2\right).
\end{gather*}
The $q$-Laguerre polynomials in equation \eqref{qLag} satisfy the orthogonality relation
\begin{gather*}
\int_0^{\frac{1}{1-q^2}}d_{q^2}u\,u^\alpha\cL_j^{(\alpha)}\big(u|q^2\big)\cL_k^{(\alpha)}\big(u|q^2\big)E_{q^2}(-u)=
\delta_{jk}q^{2(j+1)(\alpha+1+j)}\frac{\Gamma_{q^2}(j+\alpha+1)}{[j]_{q^2}!},
\end{gather*}
see \cite[equation (3.20.2)]{Koekoek}. Using the calculation rules in~\cite[Lemma 10]{MR2595267}, this can be rewritten~as
\begin{gather}
 \int_0^{\frac{1}{\sqrt{1-q}}}d_qr\, r^{2\alpha+1}\cL_j^{(\alpha)}\left(\frac{r^2}{1+q}|q^2\right)\cL_k^{(\alpha)}\left(\frac{r^2}{1+q}|q^2\right) E_{q^2}\left(\frac{-r^2}{1+q}\right)\nonumber\\
\qquad{} =\delta_{jk}\frac{q^{2(j+1)(j+\alpha+1)}\Gamma_{q^2}(j+\alpha+1)(1+q)^{\alpha}}{[j]_{q^2}!}.\label{orthLag1}
\end{gather}
One of the orthogonality relations for $q$-Laguerre polynomials in equation~\eqref{qLaginv} is
\begin{gather}
 \int_0^{\gamma\cdot\infty}d_qt\, t^{2\alpha+1}\cL_j^{(\alpha)}\left(\frac{q^2t^2}{1+q}|q^{-2}\right)\cL_k^{(\alpha)}\left(\frac{q^2t^2}{1+q}|q^{-2}\right)
 e_{q^2}\left(\frac{-q^2t^2}{1+q}\right)\nonumber\\
\qquad{} =\delta_{jk}\frac{q^{-(j+\alpha)(j+\alpha+1)}\Gamma_{q^2}(j+\alpha+1)
(1+q)^{\alpha}}{[j]_{q^2}!\,q^{(j+1)(j+2\alpha+2)}}d\left(\frac{\gamma}{\sqrt{1+q}},\alpha\right),\label{orthLag2}
\end{gather}
see \cite[equation~(3.21.3)]{Koekoek}, with
\begin{gather*}
d\left(\frac{\gamma}{\sqrt{1+q}},\alpha\right) = \frac{q^{(\alpha+1)(\alpha+2)}}{(1+q)^\alpha\Gamma_{q^2}(\alpha+1)}\int_0^{\gamma\cdot\infty}d_qt\, t^{2\alpha+1}\,e_{q^2}\left(\frac{-q^2t^2}{1+q}\right).
\end{gather*}
The function $d(\lambda,\alpha)$ therefore satisf\/ies
\begin{gather*}
d\big(\sqrt{\gamma},\alpha\big) = \frac{q^{\alpha(\alpha+1)}}{\Gamma_{q^2}(\alpha+1)}\int_0^{\gamma\cdot\infty}d_{q^2}u\,u^\alpha e_{q^2}(-u).
\end{gather*}
Partial integration implies that this function satisf\/ies $d(\lambda,\alpha+1)=d(\lambda,\alpha)$ for $\alpha>-1$. The explicit expression for $d$ can be found from \cite[equation~(3.21.3)]{Koekoek}.

\begin{remark}
The $q$-Laguerre polynomials do not form a complete orthogonal system for the Hilbert space corresponding to the measure in equation \eqref{orthLag2}. In \cite{MR1803885} the compliment of the basis is constructed. The corresponding functions derived in~\cite[Section~4]{MR1803885} (with a suitable renormalization) will therefore be annihilated by the $q$-Hankel transform $\cH_\nu^{q,\gamma}$ in the subsequent Def\/inition~\ref{defHankel}. This follows from the same calculation that leads to the subsequent equation~\eqref{preHank2}.
\end{remark}

The $q$-Gegenbauer polynomials, see \cite[equation~(2.19)]{MR1340930}, are given by
\begin{gather}
\label{qGeg}
C^\lambda_n(q;t)=\sum_{j=0}^{\lfloor\frac{n}{2}\rfloor}\frac{(-1)^jq^{j(j-1)}}{[j]_{q^2}![n-2j]_q!}
\frac{(q^{2\lambda};q^2)_{n-j}}{(1-q^2)^{n-j}}((1+q)t)^{n-2j}.
\end{gather}
They are big $q$-Jacobi polynomials on $[-1,1]$ with the two parameters equal to $\lambda-\frac{1}{2}$, see~\cite[equation~(2.26)]{MR1340930}.

For $\nu>-1$, the f\/irst and second $q$-Bessel function, introduced by Jackson, see~\cite{MR0649849, JACKSON}, are given by
\begin{gather}
\label{Bessel1}
J_\nu^{(1)}\big(x|q^2\big)=\left(\frac{x}{1+q}\right)^\nu\sum_{i=0}^\infty\frac{(-1)^i}{[i]_{q^2}!
\Gamma_{q^2}(i+\nu+1)}\left(\frac{x}{1+q}\right)^{2i} \qquad \mbox{for} \ \ |x|<\frac{1}{1-q}
\end{gather}
and
\begin{gather}
\label{Bessel2}
J_\nu^{(2)}\big(x|q^2\big)=q^{\nu^2}\left(\frac{x}{1+q}\right)^\nu\sum_{i=0}^\infty\frac{q^{2i(i+\nu)}(-1)^i}{[i]_{q^2}!
\Gamma_{q^2}(i+\nu+1)}\left(\frac{x}{1+q}\right)^{2i}.
\end{gather}
$J_\nu^{(1)}(x|q^2)$ is analytical in the area $|x|<\frac{1}{1-q}$ and $J_\nu^{(2)}(x|q^2)$ is analytical on $\mR^+$.
The f\/irst $q$-Bessel function can be analytically continued by the relation
\begin{gather*}
J_\nu^{(1)}\big(x|q^2\big) = \frac{1}{q^{\nu^2}}e_{q^2}\left(-\frac{1-q}{1+q}x^2\right)J_{\nu}^{(2)}\big(x|q^2\big),
\end{gather*}
see \cite[Exercise~1.24]{MR1052153}, which is def\/ined for all~$x$. Since $x^{-\nu}J_\nu^{(2)}(x|q^2)$ is an entire function the formula above implies that $x^{-\nu}J_\nu^{(1)}(x|q^2)$ is analytic on $\mC$ outside the poles $\{\pm i q^{-k}(1-q)^{-1}|k\in\mN\}$. Therefore $x^{-\nu}J_\nu^{(1)}(x|q^2)$ is well-def\/ined and analytic for~$x\in\mR$.

These $q$-Bessel functions are related to the $J^{(i)}_\nu(x;q)$ in \cite[(1.13) and (1.17)]{MR0649849}  or \cite[Exercise~1.24]{MR1052153} by
\begin{gather*}
J^{(1)}_\nu\big(x|q^2\big)=J^{(1)}_\nu\big({2(1-q)}x;q^2\big),\qquad J^{(1)}_\nu\big(x|q^2\big)=q^{\nu^2}J^{(1)}_\nu\big(2(1-q)x;q^2\big).
\end{gather*}

The generating functions for the $q$-Laguerre polynomials are given by
\begin{gather}
\label{BesselLag1}
J_\alpha^{(1)}\big(rt|q^2\big) = \left(\frac{rt}{1+q}\right)^{\alpha}\sum_{j=0}^\infty\frac{\cL_j^{(\alpha)}\big(\frac{r^2}{1+q}|q^2\big)}{\Gamma_{q^2}(\alpha+j+1)}
\frac{t^{2j}}{(1+q)^j}e_{q^2}\left(-\frac{q^2t^2}{1+q}\right),\\
\label{BesselLag2}
J_\alpha^{(2)}\big(qrt|q^2\big) = \left(\frac{rt}{1+q}\right)^{\alpha}\sum_{j=0}^\infty\frac{q^{(j+\alpha)(j+1+\alpha)}\cL_j^{(\alpha)}
\big(\frac{q^2t^2}{1+q}|q^{-2}\big)}{\Gamma_{q^2}(\alpha+j+1)}\frac{r^{2j}}{(1+q)^j}E_{q^2}\left(-\frac{r^2}{1+q}\right).
\end{gather}
This follows from direct calculations, they are equivalent to~\cite[formulas~(3.20.11) and (3.21.13)]{Koekoek}.

\subsection[The Howe dual pair and harmonic oscillator on $\mR^m_q$]{The Howe dual pair and harmonic oscillator on $\boldsymbol{\mR^m_q}$}

Quantum spaces are spaces where the variables have braid statistics. The commutation relations are generalizations of the bosonic or fermionic ones by an $\hat{R}$-matrix. The algebra of functions on a quantum space can be seen as the algebra $\cO$ of formal power series in non-commuting variables $x^1,\dots,x^m$,
\begin{gather*}
\cO = \mC[[x^1,\dots,x^m]]/I
\end{gather*}
with $I$ the ideal generated by the commutation relations of the variables. We consider quantum spaces which satisfy the Poincar\'e--Birkhof\/f--Witt property, which states that the dimension of the space of homogeneous polynomials of a certain degree is the same as in the commutative case. Superspaces, for instance, do not satisfy this property.

We focus on the case of the quantum Euclidean space $\mR^m_q$. The relations for the variables can e.g.\ be found in \cite{MR1328733, MR2006509, MR1361741}. We denote by $\cO_q$ the algebra of formal power series for the specif\/ic case of the quantum Euclidean space. The quantum Euclidean space can be def\/ined by the $\hat{R}$-matrix of the quantum orthogonal group $O_q(m)$, see~\cite{MR1097174, MR1239953}. The matrix $\hat{R}\in\mC^{(m\times m)\times(m\times m)}$ can be expressed in terms of its projection operators as $\hat{R}=q\cP_S-q^{-1}\cP_A+q^{1-m}\cP_1$, and is symmetric, $\hat{R}^{ij}_{kl}=\hat{R}_{ij}^{kl}$. The matrix $\hat{R}$ depends on the parameter $q$ and returns to the undeformed case when $q\to1$ ($\lim\limits_{q\to1}\hat{R}^{ij}_{kl}=\delta^j_k\delta^i_l$). The antisymmetric part def\/ines the commutation relations
\begin{gather*}
\cP_A^{ij}x\otimes x=\left(\cP_A\right)^{ij}_{kl}x^kx^l = 0.
\end{gather*}
We will always use the summation convention. The singlet part def\/ines the metric $C_{ij}=C^{ij}$, by $\left(\cP_1\right)^{ij}_{kl}=\frac{C^{ij}C_{kl}}{C}$ with $C=C^{ij}C_{ij}=(1+q^{2-m})[m/2]_{q^2}$. The metric satisf\/ies the relation
\begin{gather}
\label{CR}
C_{jl}\big(\hat{R}^{\pm 1}\big)^{lk}_{st} = \big(\hat{R}^{\mp 1}\big)^{kl}_{js}C_{lt}
\end{gather}
and is its own inverse, $C_{ij}C^{jk}=\delta^k_i$. The braid matrix also satisf\/ies the relation
\begin{gather}
\label{CR2}
C_{ij}\big(\hat{R}^{-1}\big)^{ij}_{kl} = q^{m-1}C_{kl}.
\end{gather}

The generalized norm squared is then def\/ined as $\bold{x}^2=x^iC_{ij}x^j$. This norm squared is central in the algebra $\cO_q$ and is invariant under the co-action of $O_q(m)$. The explicit expressions for the coaction of $O_q(m)$ or the dually related action of $\cU_q(\mathfrak{so}(m))$ can be found in e.g.~\cite{MR1138053, MR1239953, MR1328733, MR2006509}. In order to obtain a Fourier transform a second set of coordinates is needed, denoted by $\bold{y}$, which is a copy of the $\bold{x}$ coordinates. The commutation relations between the $\bold{x}$ and $\bold{y}$ coordinates are given by $y^ix^j=q^{-1}\hat{R}^{ij}_{kl}x^ky^l$, see~\cite{MR1303081, MR1235979, MR1317458}.

The dif\/ferential calculus on $\mR^m_q$ was developed in~\cite{MR1138053, MR1182031}, the action of the partial derivatives is determined by
\begin{gather}
\label{defpart}
\partial^ix^j = C^{ij}+q\big(\hat{R}^{-1}\big)^{ij}_{kl}x^k\partial^l .
\end{gather}
The Laplace operator on $\mR^m_q$ is given by $\Delta=\partial^iC_{ij} \partial^j$. It is central in the algebra generated by the partial derivatives and is $O_q(m)$-invariant. The commutation relations for the partial derivatives can be expressed using $\cP_A$ or as
\begin{gather}
\label{Rxx}
\hat{R}^{ij}_{kl}\partial^k\partial^l = q\partial^i\partial^j+\frac{1-q^2}{q^{m-1}(1+q^{2-m})} C^{ij}\Delta,
\end{gather}
see \cite{MR1182031}. Formulas \eqref{CR} and \eqref{CR2} yield
\begin{gather}
\label{Laplx}
\Delta x^j=\mu\partial^j+q^2x^j\Delta \qquad \mbox{and} \qquad \partial^j\bold{x}^2=\mu x^j+q^2\bold{x}^2\partial^j\qquad  \mbox{with} \quad\mu=1+q^{2-m}.
\end{gather}
The dilatation operator is given by
\begin{gather}
\label{dil}
\Lambda=1+\big(q^2-1\big)x^iC_{ij}\partial^j+\frac{(q^2-1)^2}{q^{m-2}\mu^2}\bold{x}^2\Delta
\end{gather}
and satisf\/ies $\Lambda x^i=q^2x^i\Lambda $. For $u\in\mR$, $\Lambda^u$ is def\/ined by $\Lambda^u x^i=q^{2u}x^i\Lambda^u$ and $\Lambda^u(1)=1$.

The elements of $\cO_q$ corresponding to f\/inite summations are the polynomials, the correspon\-ding algebra is denoted by $\cP$. The space $\cP_k$ is def\/ined as the space of the polynomials $P$ in $\cP$ which satisfy
\begin{gather}
\label{propdil}
\Lambda (P)=q^{2k}P \qquad \mbox{or} \qquad \frac{\Lambda-1}{q^2-1}P=[k]_{q^2}P.
\end{gather}
For $f$ analytical in the origin, $f(\bold{x}^2)\in\cO_q$ is def\/ined by the Taylor expansion of $f$. Equa\-tion~\eqref{Laplx} leads to
\begin{gather}
\label{partialr}
\partial^jf\big(\bold{x}^2\big) = x^j\mu\partial_{\bold{x}^2}^{q^2}f\big(\bold{x}^2\big)+f\big(q^2\bold{x}^2\big)\partial^j
\end{gather}
for general functions of $\bold{x}^2$ and the $q$-derivative as def\/ined in formula~\eqref{qder}. The relation{\samepage
\begin{gather}
\label{berpx2}
\partial_{\bold{x}^2}^{q^2}f\big(\bold{x}^2\big) = \left[\frac{1}{(1+q)t}\partial_t^qf\big(t^2\big)\right]_{t^2=\bold{x}^2}
\end{gather}
is a useful calculation rule.}

There exists a second dif\/ferential calculus on $\mR^m_q$ of partial derivatives $\overline{\partial^j}$, which is obtained from the unbarred one by replacing $\partial^j$, $q$, $\hat{R}$, $C$ by $\overline{\partial^j}$, $q^{-1}$, $\hat{R}^{-1}$, $C$. In particular, the relation
\begin{gather}
\label{partialrbar}
\overline{\partial^j}f\big(\bold{x}^2\big) = x^jq^{m-2}\mu\partial_{\bold{x}^2}^{q^{-2}}f\big(\bold{x}^2\big)+f\big(q^{-2}\bold{x}^2\big)\overline{\partial^j}
\end{gather}
holds. The algebra generated by the variables and partial derivatives $\partial^j$ is denoted ${\rm Dif\/f}(\mR^m_q)$. The algebra generated by the variables and partial derivatives $\overline{\partial^j}$ is the same algebra. The polynomial null-solutions of $\overline{\Delta}$ are the same as those of $\Delta$. The space of the null-solution of degree $k$ is denoted by $\cS_k$, so $\cS_k=\cP_k\cap \ker \Delta$.

\begin{definition}
The operator $E$ is given by $E=[\frac{m}{2}]_{q^2}+q^mx^iC_{ik}\partial^k$.
\end{definition}

Using the expression for the dilation operator in formula \eqref{dil} the operator $E$ can be expressed as
\begin{gather*}
E= \left[\frac{m}{2}\right]_{q^2}+q^m\left(\frac{\Lambda-1}{q^2-1}-q^{m-2}\big(q^2-1\big)\bold{x}^2\Delta\right).
\end{gather*}
Property \eqref{propdil} then implies that for $S_k\in\cS_k$,
\begin{gather}
\label{Eharm}
ES_k = \left[\frac{m}{2}+k\right]_{q^2}S_k
\end{gather}
holds. Together with $\Delta$ and $\bold{x}^2$, this operator $E$ generates an algebra which is a $q$-deformation of the universal enveloping algebra of $\mathfrak{sl}_2$.
\begin{theorem}
\label{sl2}
The operators $\Delta/\mu$, $\bold{x}^2/\mu$ and $E$ generate the quantum algebra $\cU_q(\mathfrak{sl}_2)$,
\begin{gather}
\label{commLaplxkwad}
\left[\Delta/\mu,\bold{x}^2/\mu\right]_{q^4}=E,\qquad
\left[E,\bold{x}^2/\mu\right]_{q^2}=[2]_{q^2} \bold{x}^2/\mu,\qquad
\left[\Delta /\mu,E\right]_{q^2}=[2]_{q^2} \Delta/\mu.
\end{gather}
\end{theorem}

\begin{proof}\sloppy
Combining equations \eqref{Laplx} and \eqref{CR2} yields equation \eqref{commLaplxkwad}. Equation~\eqref{Laplx} implies $x^iC_{ij}\partial^j\bold{x}^2 =\mu \bold{x}^2+q^2\bold{x}^2x^iC_{ij}\partial^j$, which leads to the second relation. The third relation is calculated in the same way.
\end{proof}

\begin{remark}
As the generators of $\cU_q(\mathfrak{sl}_2)$ are $O_q(m)$-invariant, this quantum algebra and quantum group form the Howe dual pair $(O_q(m),\cU_q(\mathfrak{sl}_2))$, or $(\cU_q(\mathfrak{so}(m)),\cU_q(\mathfrak{sl}_2))$.
\end{remark}

The quantum algebra $\cU_q(\mathfrak{sl}_2)$ is equal to the one in~\cite{MR2595267}. In \cite{MR2595267}, $\cU_q(\mathfrak{sl}_2)$ was generated by the standard Euclidean norm squared $r^2$ on $\mR^m$ and a $q$-deformation of the Laplace operator~$\Delta_q$. Since $\Delta_q$ is still $O(m)$-invariant, the Howe dual pair $(O(m),\cU_q(\mathfrak{sl}_2))$ appeared. Because the $O_q(m)$-invariant harmonic operators on quantum Euclidean space in the present paper generate the same quantum algebra we obtain an important connection between these two theories. In particular the $O_q(m)$-invariant Fourier transform developed in the current paper can be used to construct the $O(m)$-invariant $q$-Fourier transform on Euclidean space, as will be done in Section~\ref{qFeucl}.

\begin{lemma}[Fischer decomposition]
\label{Fischer}
The space $\cP$ decomposes into irreducible pieces under the action of $\cU_q(\mathfrak{so}(m))$ as $($see {\rm \cite{MR2006509, MR1328733})}
\begin{gather*}
\cP=\bigoplus_{j=0}^{\infty}\bigoplus_{k=0}^\infty\bold{x}^{2j}\cS_{k}.
\end{gather*}
\end{lemma}

The operator identities in Theorem~\ref{sl2} yield
\begin{gather*}
E\big(\bold{x}^{2l}S_k\big)=\left(\left[\frac{m}{2}+k+l\right]_{q^2}+q^2[l]_{q^2}\right)\bold{x}^{2l}S_k
\end{gather*}
and $\Delta(\bold{x}^{2l}S_k)=\mu^2[l+k+\frac{m}{2}-1]_{q^2}[l]_{q^2}\bold{x}^{2l-2}S_k$. These calculations and the previous results lead to
\begin{theorem}[Howe duality]\sloppy
The decomposition of $\cP$ into irreducible representations of $\cU_q(\mathfrak{so}(m))$ is given in Lemma~{\rm \ref{Fischer}}. Each space $\bigoplus_j \bold{x}^{2j}\cS_k$ is a lowest weight module of $\cU_q(\mathfrak{sl}_2)$ with weight vectors $\bold{x}^{2j}\cS_k$, the lowest weight vector is $\cS_k$ with weight $[m/2+k]_{q^2}$. The Fischer decomposition of~$\cP$ therefore is a multiplicity free irreducible direct sum decomposition under the joint action of $\cU_q(\mathfrak{sl}_2)\times\cU_q(\mathfrak{so}(m))$.
\end{theorem}

The antilinear involutive antihomomorphism $\ast$ on ${\rm Dif\/f}(\mR^m_q)$ is def\/ined by $(AB)^\ast=B^\ast A^\ast$, $(x^j)^\ast=x^kC_{kj}$,  $(\partial^j)^\ast=-q^{-m}\overline{\partial^k}C_{kj}$ and $\lambda^\ast=\overline{\lambda}$ with $\lambda\in\mC$ and $\overline{\cdot}$ complex conjugation. This yields
\begin{gather*}
\big(\bold{x}^2\big)^\ast=\bold{x}^2 \qquad \mbox{and}\qquad \Delta^\ast=q^{-2m}\overline{\Delta}.
\end{gather*}

The harmonic oscillator on quantum Euclidean space was studied in \cite{MR1097174, MR1288667, MR1239953}. The two Hamiltonians (with an unimportant dif\/ferent normalization compared to~\cite{MR1239953}) are given by
\begin{gather}
\label{HOs}
h=\frac{1}{2}\left(-\Delta+\bold{x}^2\right),\qquad h^\ast=\frac{1}{2}\left(-{\Delta}^\ast+\bold{x}^2\right).
\end{gather}
Both operators have the same eigenvalues.

\subsection{Integration and Fourier theory on quantum spaces}
In \cite{MR1303081} a method was prescribed to generalize $q$-integration to higher dimensions in the context of quantum spaces. Gaussian-induced integration for general $\hat{R}$-matrices is def\/ined assuming there is a matrix $\eta\in\mR^{m\times m}$ and a solution $g_{\eta}\in\cO$ of the equation
\begin{gather}
\label{Gaussian}
-\eta^i_{ j}\partial^jg_{\eta}=x^i g_\eta.
\end{gather}
Integration $\int$ on the space $\cP g_{\eta}$, with $\cP$ the polynomials on the quantum space is then uniquely def\/ined by demanding $\int\circ \partial^i =0$, $i=1,\dots,m$. For $f\in\cP$ the integral $\int f g_{\eta}$ is of the form
\begin{gather}
\label{Gaussian2}
\int fg_{\eta}=Z[f]I(g_\eta),
\end{gather}
with $I(g_\eta)=\int g_\eta$ and $Z$ a functional on~$\cP\subset\cO$. Superspace with purely bosonic and fermionic coordinates can be seen as a limit of braided spaces, typically for $q\to-1$, see e.g.~\cite{MR1413908}. From this point of view it is interesting to note that the Berezin integral can also be constructed in this setting. In \cite{DBS5, MR2539324} this led to integration over the supersphere and a new interpretation of the Berezin integral.

An explicit example of this construction was already def\/ined on quantum Euclidean space, see~\cite{MR1239953}. In~\cite{MR1408120} it was shown that this integration can be def\/ined and generalized using integration over the quantum Euclidean sphere. In Section~\ref{DefFourier} we will show how this approach follows from harmonic analysis on quantum Euclidean space.

In \cite{MR1303081} a general procedure to construct a Fourier transform on quantum spaces was developed. First the appropriate Gaussian-induced integration $\int$ should be constructed and the exponential or Fourier kernel (see \cite{MR1235979, MR2099760}) calculated. The Fourier transform on a braided-Hopf algebra $B$ with left dual Hopf algebra $B^\ast$  is a map $\cF:B\to B^\ast$. The co-ordinates for $B$ are denoted by $\bold{x}$ and for $B^\ast$ by $\bold{y}$, the Fourier transforms are given by
\begin{gather*}
\cF[f(\bold{x})](\bold{y})=\int_x f(\bold{x})\exp_{\hat{R}}(\bold{x}|\bold{y}), \qquad \cF^\ast[f(\bold{y}](\bold{x})=\int^\ast_y f(\bold{y})\exp_{\hat{R}}(\bold{x}|\bold{y}).
\end{gather*}
These Fourier transforms are each others inverse, $\cF^\ast\cF=\mbox{Vol}\, S$, with $S$ the antipode on $B$. As an explicit example we consider the braided line $B=\mC[x]$, with braided-Hopf algebra structure as introduced in \cite{MR1182915} given by
\begin{gather*}
\Delta x^k=\sum_{j=0}^k\frac{[k]_q!}{[k-j]_q![j]_q!}x^j\otimes x^{k-j},\qquad Sx^k=(-1)^kq^{\frac{k(k-1)}{2}}x^k,\qquad\epsilon x^k=\delta_{k0}.
\end{gather*}
The dually-paired Hopf algebra $B^\ast$ is the same Hopf algebra with variable~$y$. The relation $xy=qyx$ holds. The exponential is $\exp(x|y)=\sum\limits_{k=0}^\infty\frac{x^ky^k}{[k]_q!}$ and satisf\/ies $\partial_x^q\exp(x|y)=\exp(x|y)y$. The Fourier transforms take the form
\begin{gather}
\label{Fourline}
\cF[f](y)=\int_{-\gamma\cdot\infty}^{\gamma\cdot\infty}d_qxf(x)\exp(x|y) \qquad \mbox{and}\qquad \cF^\ast[f](x)=\int_{-\delta\cdot\infty}^{\delta\cdot\infty}d_qyf(y)\exp(x|y).
\end{gather}
The theory of \cite{MR1303081} then implies $\cF^\ast\cF[f](x)=Sf(x)\mbox{Vol}_{\gamma,\delta}$. We can rewrite this in a way that will be more closely related to our approach of the Fourier transform on $\mR^m_q$. Def\/ine $g(y)=\frac{1}{{\rm Vol}_{\gamma,\delta}}\cF[f](y)$, then the def\/inition of the antipode implies
\begin{gather*}
f(-x) = \int_{-\delta\cdot\infty}^{\delta\cdot\infty}d_qyg(y)\sum_{k=0}^\infty\frac{q^{\frac{k(k+1)}{2}}y^kx^k}{[k]_q!}.
\end{gather*}
In this equation and in the f\/irst equation of~\eqref{Fourline}, there are no coordinates which have to be switched before integration. This implies that we can assume $x$ and $y$ commute and write the equations above as
\begin{gather*}
g(y)=\frac{1}{{\rm Vol}_{\gamma,\delta}}\int_{-\gamma\cdot\infty}^{\gamma\cdot\infty}d_qx\,f(x)\,e_q(xy)\qquad \mbox{and}\qquad f(x)=\int_{-\delta\cdot\infty}^{\delta\cdot\infty}d_qy\,g(y)\,E_q(-qyx).
\end{gather*}

In \cite{MR1448685} a closely related analytical approach was given to the one dimensional Fourier transform above. Consider real commuting variables~$x$ and~$y$. Using the orthogonality relations and the generating function of the Hermite polynomials in equation~\eqref{Hermite}, it is possible to prove
\begin{gather}
\frac{1}{2\Gamma_{q^2}(\frac{1}{2})}\int_{-\frac{1}{\sqrt{1-q}}}^{\frac{1}{\sqrt{1-q}}}d_qx\, \left[H_k^q\left(\frac{x}{\sqrt{1+q}}\right)E_{q^2}\left(-\frac{x^2}{1+q}\right)\right]e_q(-ixy)\nonumber\\
\qquad{}
=\frac{(q+1)^{\frac{k-1}{2}}}{i^k}q^{\frac{1}{2}(k+1)(k+2)}y^ke_{q^2}\left(-\frac{q^2y^2}{1+q}\right),\label{1dFour}
\end{gather}
which is equivalent to~\cite[equation~(8.7)]{MR1448685}. For every $\delta\in\mR^+$,
\begin{gather}
\frac{1}{C_\delta}\int_{-\delta\cdot\infty}^{\delta\cdot\infty}d_qy\, \left[y^ke_{q^2}\left(-\frac{q^2y^2}{1+q}\right)\right]E_q(iqyx)\nonumber\\
\qquad =\frac{i^k}{q^{\frac{1}{2}(k+1)(k+2)}(1+q)^{\frac{k-1}{2}}}H_k^q\left(\frac{x}{\sqrt{1+q}}\right)E_{q^2}\left(-\frac{x^2}{1+q}\right)\label{1dFourb}
\end{gather}
holds for $C_\delta$ some constant depending on $\delta$, see~\cite[equation~(8.21)]{MR1448685}. So the two Fourier transforms as def\/ined in equations~\eqref{1dFour} and~\eqref{1dFourb} can be regarded as each others inverse, which was to be expected from the theory of~\cite{MR1303081}. There is however one dif\/ference between the explicit Fourier transform in~\cite{MR1448685} and the abstract theory in~\cite{MR1303081}. While the inverse Fourier transform remains unchanged, the integration for the Fourier transform is limited to a f\/inite interval. This will be explained in the subsequent Lemma~\ref{intexpinf}. However, using property~\eqref{intinfeind} the integral can be replaced by $\int_{-\frac{1}{\sqrt{1-q}}\cdot\infty}^{\frac{1}{\sqrt{1-q}}\cdot\infty}$. So the analytical approach of~\cite{MR1448685} recovers the theory from~\cite{MR1303081} with an imposed limitation on~$\gamma$. For other~$\gamma$, the theory from~\cite{MR1303081} would still hold, but the constant ${\rm Vol}_{\gamma,\delta}$ is inf\/inite.

\section[The $q$-Hankel transforms]{The $\boldsymbol{q}$-Hankel transforms}
\label{Hanksect}

In this section we def\/ine two $q$-Hankel transforms using the f\/irst and second $q$-Bessel function. These transforms will act as each others inverse. This is a generalization of the result in formulas~\eqref{1dFour} and~\eqref{1dFourb}. By evaluating the Fourier transform on the appropriate functions in Section~\ref{boch} the f\/irst of these $q$-Hankel transforms will appear.

We will calculate the $q$-Hankel transforms of the $q$-Laguerre and little $q$-Laguerre polynomials weighted with a $q$-Gaussian. The undeformed Fourier--Gauss transform of these polynomials was already studied in \cite{MR1479780}. There also exists a third $q$-Bessel function besides the ones in equations~\eqref{Bessel1} and~\eqref{Bessel2}. We will not explicitly need the third type, but it is interesting to note that in~\cite{MR1069750, MR2445269} the corresponding $q$-Hankel transforms were constructed. These Hankel transforms could also be used to def\/ine an $O_q(m)$-invariant Fourier transform on $\mR^m_q$. This would have the advantage that the Fourier transform is its own inverse. That Fourier transform would however not behave well with respect to the derivatives on $\mR^m_q$. This is already the case for the Fourier transform on the braided line, as is proven in~\cite{MR1069750}. The braided line corresponds to $\mR^m_q$ for $m=1$.

In anticipation of the connection with the Fourier transform on quantum Euclidean space we will scale the $q$-Hankel transforms in the following def\/inition with~$\mu$, see equation~\eqref{Laplx}, although at this stage any constant could be used. The reason for the appearance of unf\/ixed constants~$\beta$,~$\gamma$ will become apparent in Sections~\ref{boch}, \ref{behave} and \ref{FourHO}.
\begin{definition}
\label{defHankel}
For $\beta,\gamma\in\mR^+$ and $\nu\ge-\frac{1}{2}$, the $q$-Hankel transforms are given by
\begin{gather*}
\overline{\cH}_{\nu}^{q,\beta}[f(r)](t) = \frac{1+q}{\mu}\int_0^{\sqrt{\frac{\mu}{(1-q^2)\beta}}}d_qr\frac{J_\nu^{(1)}\big(\frac{1+q}{\mu}rt|q^2\big)}{(rt)^{\nu}}r^{2\nu+1}[f(r)]
\end{gather*}
and
\begin{gather*}
\cH^{q,\gamma}_{\nu}[f(t)](r) = \frac{1+q}{\mu}\int_0^{\gamma\cdot\infty}d_qt\frac{J_\nu^{(2)}\big(q\frac{1+q}{\mu}rt|q^2\big)}{(rt)^{\nu}}t^{2\nu+1}[f(t)].
\end{gather*}
\end{definition}

In this def\/inition it is not specif\/ied on which function spaces the $q$-Hankel transforms act. At the moment we def\/ine them on functions for which the expression exists.

In order to connect the Fourier transform on $\mR^m_q$ with these $q$-Hankel transforms we def\/ine the following transformations,
\begin{gather*}
\overline{\cF_{\nu}}^{q,\beta}[\psi]\big(t^2\big)=\overline{\cH}_{\nu}^{q,\beta}[\psi\circ\Upsilon](t)
\qquad \mbox{and} \qquad \cF_{\nu}^{q,\gamma}[\psi]\big(r^2\big)=\cH_{\nu}^{q,\gamma}[\psi\circ\Upsilon](r),
\end{gather*}
with $\Upsilon(u)=u^2$.

In order to prove the properties of the $q$-Hankel transforms we will need some identities of the $q$-Bessel functions in equations~\eqref{Bessel1} and~\eqref{Bessel2}. These are summarized in the following lemma.

\begin{lemma}
\label{lemBessel1}
The first and second $q$-Bessel functions satisfy
\begin{alignat*}{3}
& (i)& &\partial_u^q\frac{J_{\nu}^{(1)}(u|q^2)}{u^\nu}=-u\frac{J_{\nu+1}^{(1)}(u|q^2)}{u^{\nu+1}},\qquad \partial_u^{q^{-1}}\frac{J_{\nu}^{(2)}(qu|q^2)}{u^\nu}=-qu\frac{J_{\nu+1}^{(2)}(qu|q^2)}{u^{\nu+1}}, &\\
& (ii)& &\frac{J_{\nu+1}^{(1)}(u|q^2)+J_{\nu-1}^{(1)}(u|q^2)}{u^{\nu-1}}=[2\nu]_q\frac{J_{\nu}^{(1)}(qu)}{(qu)^\nu},& \\ &&& \frac{J_{\nu+1}^{(2)}(qu|q^2)+J_{\nu-1}^{(2)}(qu|q^2)}{u^{\nu-1}}=[2\nu]_q\frac{J_{\nu}^{(2)}(u)}{(qu)^\nu},&\\
& (iii) \ \ & &\partial_u^q J_{\nu}^{(1)}(u|q^2)u^\nu=J_{\nu-1}^{(1)}(u|q^2)u^\nu\qquad
 \mbox{and}\qquad \partial_u^q J_{\nu}^{(2)}(u|q^2)u^\nu=q^\nu J_{\nu-1}^{(2)}(qu|q^2)u^\nu.&
\end{alignat*}
\end{lemma}

\begin{proof}
The right-hand side of the second property (for the f\/irst $q$-Bessel function) can be calculated using $[\nu]_{q^2}q^{2i}=[\nu+i]_{q^2}-[i]_{q^2}$,
\begin{gather*}
[\nu]_{q^2}(1+q)\frac{J_\nu^{(1)}(qu)}{(qu)^\nu} \\
\qquad{}
=\frac{1}{(1+q)^{\nu-1}}\sum_{i=0}^\infty\left(\frac{(-1)^i}{[i]_{q^2}!\Gamma(i+\nu)}\left(\frac{u}{1+q}\right)^{2i}
-\frac{(-1)^i}{[i-1]_{q^2}!\Gamma(i+\nu+1)}\left(\frac{u}{1+q}\right)^{2i}\right)\\
\qquad{} =\frac{J_{\nu-1}^{(1)}(u|q^2)+J_{\nu+1}^{(1)}(u|q^2)}{u^{\nu-1}}.
\end{gather*}
The f\/irst and the third property follow from a direct calculation. The left-hand sides of the properties can also be obtained from \cite[Exercise~1.25]{MR1052153}.
\end{proof}

\begin{corollary}
\label{lemBesselextra}
The second $q$-Bessel functions satisfy the following relation:
\begin{gather*}
\partial_u^q u^{\nu+1}J_{\nu-1}^{(2)}\big(u|q^2\big) = [2\nu]_qu^\nu J_{\nu-1}^{(2)}\big(u|q^2\big)-q^{\nu+1}u^{\nu+1}J_{\nu}^{(2)}\big(qu|q^2\big).
\end{gather*}
\end{corollary}
\begin{proof}
This is a direct consequence of the second formula in Lemma~\ref{lemBessel1}$(i)$.
\end{proof}

Combining generating function \eqref{BesselLag1} and orthogonality relation \eqref{orthLag1} yields
\begin{gather}
\int_0^{\frac{1}{\sqrt{1-q}}}d_qr \frac{J_\nu^{(1)}(rt|q^2)}{(rt)^{\nu}}r^{2\nu+1}\left[\cL_j^{(\nu)}\left(\frac{r^2}{1+q}|q^2\right)E_{q^2}\left(\frac{-r^2}{1+q}\right)\right]
\nonumber\\
\qquad {} =\frac{q^{2(j+1)(j+\nu+1)}}{[j]_{q^2}!}\frac{t^{2j}}{(1+q)^j}e_{q^2}\left(-\frac{q^2t^2}{1+q}\right).\label{preHank1}
\end{gather}
Generating function \eqref{BesselLag2} and orthogonality relation \eqref{orthLag2} lead to
\begin{gather} \int_0^{\gamma\cdot\infty}d_qt\frac{J_\nu^{(2)}(qrt|q^2)}{(rt)^{\nu}}t^{2\nu+1}\left[\cL_j^{(\nu)}
\left(\frac{q^2t^2}{1+q}|q^{-2}\right)e_{q^2}\left(\frac{-q^2t^2}{1+q}\right)\right]\nonumber\\
\qquad {}= d\left(\frac{\gamma}{\sqrt{1+q}},\nu\right)\frac{q^{-(j+1)(j+2\nu+2)}}{[j]_{q^2}!}\frac{r^{2j}}{(1+q)^j}E_{q^2}
\left(-\frac{r^2}{1+q}\right).\label{preHank2}
\end{gather}

The following expansion of a monomial in terms of the $q$-Laguerre polynomials is a direct consequence of equation \eqref{BesselLag2},
\begin{gather*}
\frac{t^{2j}}{(1+q)^j[j]_{q^2}!}=\sum_{i=0}^j\frac{(-1)^i(q^{2i+2\nu+2};q^2)_{(j-i)}}{[j-i]_{q^2}!(1-q^2)^{j-i}}
\frac{q^{(j-i)(j-i+1)+(i+1)(i+2\nu+2)}}{q^{2(j+1)(j+\nu+1)}}\cL_i^{(\nu)}\left(\frac{q^2t^2}{1+q}|q^{-2}\right).
\end{gather*}
Applying this yields
\begin{gather*}
\int_0^{\gamma\cdot\infty}d_qt\frac{J_\nu^{(2)}(qrt|q^2)}{(rt)^{\nu}}t^{2\nu+1}
\left[\frac{t^{2j}}{(1+q)^j[j]_{q^2}!}e_{q^2}\left(\frac{-q^2t^2}{1+q}\right)\right]\\
\quad{}=\sum_{i=0}^j\frac{(-1)^i(q^{2i+2\nu+2};q^2)_{(j-i)}}{[j-i]_{q^2}!(1-q^2)^{j-i}}\frac{q^{(j-i)(j-i+1)}}{q^{2(j+1)(j+\nu+1)}}
d\left(\frac{\gamma}{\sqrt{1+q}},\nu\right)\frac{1}{[i]_{q^2}!}\frac{r^{2i}}{(1+q)^i}E_{q^2}\left(-\frac{r^2}{1+q}\right)\\
\quad{}=d\left(\frac{\gamma}{\sqrt{1+q}},\nu\right) q^{-2(j+1)(j+\nu+1)}\cL^{(\nu)}_j\left(\frac{r^2}{1+q}|q^2\right)E_{q^2}\left(-\frac{r^2}{1+q}\right).
\end{gather*}
These calculations imply the following relations for the $q$-Hankel transforms in Def\/inition~\ref{defHankel}:
\begin{gather*}
\overline{\cH}_{\nu}^{q,\beta}\left[\cL_j^{(\nu)}\left(\beta\frac{r^2}{\mu}|q^2\right)E_{q^2}\left(-\beta\frac{r^2}{\mu}\right)\right](t)=
\frac{1}{\beta^{\nu+1+j}}C_jt^{2j}e_{q^2}\left(-\frac{q^2t^2}{\mu\beta}\right)\qquad\mbox{for} \ \ \beta\in\mR^+,\\
\cH^{q,\gamma}_{\nu}\left[C_jt^{2j}e_{q^2}\left(-\alpha\frac{q^2t^2}{\mu}\right)\right](r)= \frac{d\big(\frac{\sqrt{\alpha}\gamma}{\sqrt{\mu}},\nu\big) }{\alpha^{\nu+1+j}}\cL^{(\nu)}_j\left(\frac{r^2}{\alpha\mu}|q^2\right)E_{q^2}\left(-\frac{r^2}{\alpha\mu}\right)\!\qquad\mbox{for} \ \ \alpha,\gamma\in\mR^+,
\end{gather*}
with $C_j=\frac{q^{2(j+1)(j+\nu+1)}}{[j]_{q^2}!\mu^j}$.

By considering the case $\beta=1/\alpha$ we obtain
\begin{theorem}
\label{Hankelinv}
For each $\alpha,\gamma\in\mR^+$, the inverse of the $q$-Hankel transform $\cH_\nu^{q,\gamma}$ acting on $\mR[t^2]\otimes e_{q^2}(-\alpha\frac{q^2t^2}{\mu})$,
\begin{gather*}
\cH_\nu^{q,\gamma}: \ \ \mR\big[t^2\big]\otimes e_{q^2}\left(-\alpha\frac{q^2t^2}{\mu}\right) \to\mR\big[r^2\big]\otimes E_{q^2}\left(-\frac{r^2}{\alpha\mu}\right)
\end{gather*}
is given by
\begin{gather*}
\frac{1}{d\big(\frac{\sqrt\alpha \gamma}{\sqrt\mu},\nu\big)}\overline{\cH}^{q,1/\alpha}_{\nu}: \ \ \mR\big[r^2\big]\otimes E_{q^2}\left(-\frac{r^2}{\alpha\mu}\right)\to \mR\big[t^2\big]\otimes e_{q^2}\left(-\alpha\frac{q^2t^2}{\mu}\right).
\end{gather*}
\end{theorem}
This theorem for $\nu=-\frac{1}{2}$ and $\nu=\frac{1}{2}$ is identical to the results in equations \eqref{1dFour} and \eqref{1dFourb}.

In order to prove the behavior of the Fourier transform on $\mR^m_q$ we need the following properties of the $q$-Hankel transforms. The exact function spaces on which they act is again not specif\/ied, the properties hold if all the terms are well-def\/ined. In particular these lemmata hold for the function spaces in Theorem~\ref{Hankelinv}.

\begin{lemma}
\label{eigHankF}
The first $q$-Hankel transform satisfies the following properties:
\begin{alignat*}{3}
& (i)& & t^2\overline{\cH}^{q,\beta}_{\nu+1}[f(r)](t)+\overline{\cH}^{q,\beta}_{\nu-1}\big[r^2f(r)\big](t)=\mu[\nu]_q\overline{\cH}_{\nu}^{q,\beta}[f(r)](qt),& \\
& (ii)& &  \partial_t^q\overline{\cH}_{\nu}^{q,\beta}[f](t)=-\frac{1+q}{\mu}t\overline{\cH}_{\nu+1}^{q,\beta}[f](t)\qquad \mbox{and}& \\
& (iii) \quad & & \overline{\cH}_\nu^{q,\beta}\left[\frac{1}{r}\partial_r^{q^{-1}}f\right](t)=-q\frac{1+q}{\mu}\overline{\cH}_{\nu-1}^{q,\beta}[f](t)
\qquad\mbox{if}\quad f\left(q^{-1}\sqrt{\frac{\mu}{(1-q^2)\beta}}\right)=0.&
\end{alignat*}
The second $q$-Hankel transform satisfies the following properties:
\begin{alignat*}{3}
& (i)& &  r^2\cH^{q,\gamma}_{\nu+1}[f(t)](r)+\cH^{q,\gamma}_{\nu-1}\big[t^2f(t)\big](r)
=\frac{\mu[\nu]_q}{q^{2\nu}}\cH_{\nu}^{q,\gamma}[f(t)]\big(q^{-1}r\big),\\
& (ii)& & \partial_r^{q^{-1}}\cH_{\nu}^{q,\gamma}[f](r)=-q\frac{1+q}{\mu}r\cH_{\nu+1}^{q,\gamma}[f](r)\qquad \mbox{and}& \\
& (iii)\quad & & \cH_\nu^{q,\gamma}\left[\frac{1}{t}\partial_t^{q}f\right](r)=-\frac{1+q}{\mu}\cH_{\nu-1}^{q,\gamma}[f](r).
\end{alignat*}
\end{lemma}

\begin{proof}
The f\/irst property is a direct consequence of Lemma~\ref{lemBessel1}$(ii)$, the second a direct consequence of Lemma~\ref{lemBessel1}$(i)$. Property $(iii)$ is calculated using formulas~\eqref{Leibniz} and~\eqref{partint1} and Lem\-ma~\ref{lemBessel1}$(iii)$	
\begin{gather*}
\overline{ \cH}_\nu^{q,\beta}\left[\frac{1}{r}\partial_r^{q^{-1}}f\right](t)=
\int_0^{\sqrt{\frac{\mu}{(1-q^2)\beta}}}d_qr\frac{J_\nu^{(1)}\big(\frac{1+q}{\mu}rt|q^2\big)}{t^{\nu}}r^{\nu}\partial_r^{q^{-1}}f(r)\\
\hphantom{\overline{ \cH}_\nu^{q,\beta}\left[\frac{1}{r}\partial_r^{q^{-1}}f\right](t)}{} =\int_0^{\sqrt{\frac{\mu}{(1-q^2)\beta}}}d_qr\frac{J_\nu^{(1)}\big(\frac{1+q}{\mu}rt|q^2\big)}{t^{\nu}}r^{\nu}\partial_r^{q}f\big(q^{-1}r\big)q\\
\hphantom{\overline{ \cH}_\nu^{q,\beta}\left[\frac{1}{r}\partial_r^{q^{-1}}f\right](t)}{}
=-q\int_0^{\sqrt{\frac{\mu}{(1-q^2)\beta}}}d_qr\left[\partial_r^q\frac{J_\nu^{(1)}\big(\frac{1+q}{\mu}rt|q^2\big)}{t^{\nu}}r^{\nu}\right]f(r)\\
\hphantom{\overline{ \cH}_\nu^{q,\beta}\left[\frac{1}{r}\partial_r^{q^{-1}}f\right](t)}{}
=-q\frac{1+q}{\mu}\int_0^{\sqrt{\frac{\mu}{(1-q^2)\beta}}}d_qr\frac{J_{\nu-1}^{(1)}\big(\frac{1+q}{\mu}rt|q^2\big)}{t^{\nu-1}}r^{\nu}f(r).
\end{gather*}
Property $(iii)$ for the second $q$-Hankel transform is calculated similarly.
\end{proof}

\begin{lemma}
\label{lemHankelextra}
The relation
\begin{gather*}
\cH_{\nu-1}^{q,\gamma}\left[t\partial_t^qf(t)\right](r)=
\frac{1+q}{\mu}r^2\cH_\nu^{q,\gamma}[f](r)-\frac{[2\nu]_q}{q^{2\nu}}\cH_{\nu-1}^{q,\gamma}[f]\big(q^{-1}r\big)
\end{gather*}
holds for the second $q$-Hankel transform.
\end{lemma}

\begin{proof}
The left-hand side is calculated using Corollary~\ref{lemBesselextra},
\begin{gather*}
\cH_{\nu-1}^{q,\gamma}\left[t\partial_t^qf(t)\right](r)=-\frac{1+q}{\mu }\frac{1}{q^{\nu+1}}\int_0^{\gamma\cdot\infty}d_qt\partial_t^q\left[\frac{J_{\nu-1}^{(2)}\big(\frac{1+q}{\mu}rt|q^2\big)
t^{\nu+1}}{r^{\nu-1}}\right]f(t)\\
\qquad{}
 = \int_0^{\gamma\cdot\infty}d_qt\left[r^2\frac{J_\nu^{(2)}\big(q\frac{1+q}{\mu}rt|q^2\big)t^{\nu+1}}{r^\nu}f(t)-\frac{1+q}{\mu }\frac{[2\nu]_q}{q^{\nu+1}}\frac{J_{\nu-1}^{(2)}\big(\frac{1+q}{\mu}rt\big)t^{\nu}}{r^{\nu-1}}f(t)\right],
\end{gather*}
which proves the lemma.
\end{proof}

\section[Integration and Fourier transform on $\mR^m_q$]{Integration and Fourier transform on $\boldsymbol{\mR^m_q}$}

\subsection[Integration over the quantum sphere and induced integration on $\mR^m_q$]{Integration over the quantum sphere and induced integration on $\boldsymbol{\mR^m_q}$}
\label{DefFourier}

First we show how the Howe dual pair $(O_q(m),\cU_q(\mathfrak{sl}_2))$ uniquely characterizes the integration over the quantum sphere from~\cite{MR1408120}.

\begin{theorem}
\label{Pizzth}
The unique $($up to a multiplicative constant$)$ linear functional on $\cP$ invariant under the co-action of $O_q(m)$ and satisfying $\int_{\mS^{m-1}_q}\bold{x}^2R=\int_{\mS^{m-1}_q}R$ is given by the Pizzetti formula
\begin{gather*}
\int_{\mS^{m-1}_q}R= \sum_{k=0}^\infty\frac{2\left(\Gamma_{q^2}(\frac{1}{2})\right)^{m}}{\mu^{2k}[k]_{q^2}!
\Gamma_{q^2}(k+\frac{m}{2})}\big(\Delta^kR\big)(0)\qquad\mbox{for}\quad R\in\cP.
\end{gather*}
\end{theorem}

\begin{proof}
The Fischer decomposition in Lemma~\ref{Fischer} implies that the integration on~$\cP$ is uniquely determined if it is determined on each of the blocks $\bold{x}^{2l}\cS_k$. Since these blocks are irreducible $\cU_q(\mathfrak{so}(m))$-representations (or irreducible $O_q(m)$-corepresentations) the integration should be zero on each such block which is not one dimensional. This implies that the integration can only have non-zero values on the elements~$\bold{x}^{2l}$. The second property then implies $\int_{\mS^{m-1}_q}\bold{x}^{2l}=\int_{\mS^{m-1}_q}1$, which shows that the integration is uniquely determined up to the constant~$\int_{\mS^{m-1}_q}1$. It is easily checked that the Pizzetti formula satisf\/ies the conditions.
\end{proof}

We chose the normalization such that $\int_{\mS^{m-1}_q}1
=\frac{2\left(\Gamma_{q^2}(\frac{1}{2})\right)^{m}}{\Gamma_{q^2}(\frac{m}{2})}$.
This quantum sphere integration can be expressed symbolically using the f\/irst $q$-Bessel-function~\eqref{Bessel1},
\begin{gather}
\label{PizzBessel}
\int_{\mS^{m-1}_q}R = 2\left(\Gamma_{q^2}\left(\frac{1}{2}\right)\right)^{m}\left(\frac{J^{(1)}_{\frac{m}{2}-1}
\left(\frac{1+q}{\mu}\sqrt{-\Delta}|q^2\right)}{\left(\sqrt{-\Delta}/\mu\right)^{\frac{m}{2}-1}}R\right)(0).
\end{gather}

The following lemma will be important for the sequel.
\begin{lemma}\label{xS}
The Fischer decomposition $($see Lemma~{\rm \ref{Fischer})} of $x^jS_k(\bold{x})$ with $S_k\in\cS_k$ is given by
\begin{gather*}
x^jS_k=\left(x^jS_k-\bold{x}^2\frac{1}{\mu[k+m/2-1]_{q^2}}\partial^jS_k\right)+\bold{x}^2\left(\frac{1}{\mu[k+m/2-1]_{q^2}}\partial^jS_k\right)
\end{gather*}
with
\begin{gather*}
\left(x^jS_k-\bold{x}^2\frac{1}{\mu[k+m/2-1]_{q^2}}\partial^jS_k\right)\in\cS_{k+1} \!\!\!\qquad \mbox{and} \qquad\!\!\! \left(\frac{1}{\mu[k+m/2-1]_{q^2}}\partial^jS_k\right)\in\cS_{k-1}.
\end{gather*}
\end{lemma}
\begin{proof}\sloppy
Since $\Delta$ and $\partial^j$ commute $\left(\frac{1}{\mu[k+m/2-1]_{q^2}}\partial^jS_k\right)\in\cS_{k-1}$. Equations \eqref{Laplx}, \eqref{commLaplxkwad} and~\eqref{Eharm} yield
\begin{gather*}
\Delta\left(x^jS_k-\bold{x}^2\frac{1}{\mu[k+m/2-1]_{q^2}}\partial^jS_k\right)=
\mu\partial^jS_k-\mu^2E\frac{1}{\mu[k+m/2-1]_{q^2}}\partial^jS_k=0.
\end{gather*}
This can also be calculated by using the projection operator from $\cP_k$ onto $\cS_k$ developed \linebreak in~\cite{MR2006509}.
\end{proof}

Before we use the quantum sphere integration to construct integration on $\mR^m_q$ along the lines of~\cite{MR1303081} and~\cite{MR1239953}, the following lemma should be considered.
\begin{lemma}
\label{intexpinf}
For the $q$-integration in equation \eqref{gammaint} with $\gamma\in\mR^+$, the expression
\begin{gather*}
\int_0^{\gamma\cdot\infty}E_q(-t)d_qt
\end{gather*}
is infinite unless $\gamma=q^{j}\frac{1}{1-q}$ for some $j\in\mZ$. In that case it reduces to $\int_0^{\frac{1}{1-q}}E_q(-t)d_qt$.
\end{lemma}

\begin{proof}
Property \eqref{aflexp} can be rewritten as $E_q(t)=E_q(qt)\left[1+(1-q)t\right]$, which implies
\begin{gather*}
E_q\big({-}q^{-k}\gamma\big) = E_q\big({-}q^{1-k}\gamma\big)\big[1-(1-q)q^{-k}\gamma\big].
\end{gather*}

First we assume $E_q(-q^{1-k}\gamma)$ is never zero for $k\in\mZ$. The equation above then yields $\left|E_q(-q^{-k}\gamma)\right|>\left|E_q(-q^{1-k}\gamma)\right|q$ and ${\rm Sign}(E_q(-q^{-k}\gamma))\not={\rm Sign}(E_q(-q^{1-k}\gamma))$ for $k\ge N\in\mN$ with $N>\frac{\ln\left(\frac{1+q}{\gamma(1-q)}\right)}{\ln(1/q)}$. This implies that $\sum\limits_{k=N}^\infty E_q(-q^{-k}\gamma)q^{-k}$ will not converge, therefore the integral~\eqref{gammaint} will not exist.

If $E_q(-q^{1-k}\gamma)=0$ holds for some $k\in\mZ$, the unicity of the zeroes in equation~\eqref{zeroes} implies that $\gamma=\frac{q^{k-1-l}}{1-q}$ for some $k\in\mZ$ and $l\in\mN$, or $\gamma=\frac{q^{j}}{1-q}$ for some $j\in\mZ$. The integration then reduces to the proposed expression for the same reason as in equation~\eqref{intinfeind}.
\end{proof}

Now we can construct integration on $\mR^m_q$ from the quantum sphere integration according to the principle of Gaussian-induced integration. The two dif\/ferential calculi lead to dif\/ferent Gaussians and integrations. For constants $\alpha,\beta\in\mR^+$, the relations
\begin{gather*}
\partial^je_{q^2}\big({-}\alpha\bold{x}^2\big)=-\alpha\mu x^je_{q^2}\big({-}\alpha\bold{x}^2\big)\qquad \mbox{and}\qquad \overline{\partial^j} E_{q^2}\big({-}\beta\bold{x}^2\big)=-\beta q^{m-2}\mu x^jE_{q^2}\big({-}\beta\bold{x}^2\big)
\end{gather*}
imply that $e_{q^2}(-\alpha\bold{x}^2)$ and $E_{q^2}(-\beta\bold{x}^2)$ are Gaussians according to equation~\eqref{Gaussian} for the unbarred and barred calculus respectively. Def\/ine the following two integrations for $\gamma,\lambda\in\mR^+$ on polynomials weighted with undetermined radial functions,
\begin{gather*}
\int_{\gamma\cdot\mR^m_q}R_{k}(\bold{x})f\big(\bold{x}^2\big) = \int_0^{\gamma\cdot\infty}d_qr\,r^{m+k-1}f\big(r^2\big)\int_{\mS^{m-1}_q}R_{k}(\bold{x})\qquad \mbox{and}\\
\int_{\mB^m_q(\lambda)}R_{k}(\bold{x})f\big(\bold{x}^2\big) = \int_0^{\lambda}d_qr\,r^{m+k-1}f\big(r^2\big)\int_{\mS^{m-1}_q}R_{k}(\bold{x}).
\end{gather*}
The Taylor expansion of the function $f$ in the origin is assumed to converge on $\mR^+$ for the f\/irst integration and on $[0,\lambda]$ for the second integration. These integrations correspond to the Gaussian-induced integrations.
\begin{theorem}
\label{2indint}
The integrations defined above satisfy
\begin{gather*}
\int_{\gamma\cdot\mR^m_q}\partial^j=0 \qquad \mbox{on}\quad  \cP\otimes e_{q^2}\big({-}\alpha\bold{x}^2\big)\quad \mbox{for}\quad \alpha,\gamma\in\mR^+\qquad\mbox{and}\\
\int_{\mB^m_q\big(\frac{1}{\sqrt{(1-q^2)\beta}}\big)}\overline{\partial^j}=0\qquad \mbox{on}\quad \cP\otimes E_{q^2}\big({-}\beta\bold{x}^2\big)\quad\mbox{for}\quad\beta\in\mR^+.
\end{gather*}
\end{theorem}
\begin{proof}
The f\/irst property is well-known, see e.g.~\cite{MR1239953, MR1408120}. In order to prove the second property we consider the expression
\begin{gather*}
\int_{\mB^m_q(\lambda)}\overline{\partial^j}f\big(\bold{x}^2\big)S_k(\bold{x})
=q^{m-2}\mu\int_{\mB^m_q(\lambda)}\big(\partial_{\bold{x}^2}^{q^{-2}}f\big(\bold{x}^2\big)\big)x^j S_k(\bold{x})+\int_{\mB^m_q(\lambda)}f\big(q^{-2}\bold{x}^2\big)\overline{\partial^j}S_k(\bold{x}).
\end{gather*}
for a spherical harmonic $S_k\in\cS_k$. If $k\not=1$ the right-hand side is always zero, because of Lemma~\ref{xS} and the expression for $\int_{\mS^{m-1}_q}$ in Theorem~\ref{Pizzth}. In case $k=1$ the spherical harmonics are the monomials $x^i$, $i=1,\dots,m$. Therefore we calculate using equation~\eqref{berpx2}, Leibniz rule~\eqref{Leibniz} and equation~\eqref{partint1}
\begin{gather*}
\int_{\mB^m_q(\lambda)}\overline{\partial^j}f\big(\bold{x}^2\big)x^i
=q^{m-2}\mu\int_{\mB^m_q(\lambda)}\big(\partial_{\bold{x}^2}^{q^{-2}}f\big(\bold{x}^2\big)\big)x^j x^i+\int_{\mB^m_q(\lambda)}f\big(q^{-2}\bold{x}^2\big)C^{ji}\\
=\frac{2\left(\Gamma_{q^2}(\frac{1}{2})\right)^{m}C^{ji}}{\Gamma_{q^2}(1+\frac{m}{2})}\left[q^{m-2}\!\int_0^\lambda \!d_qr r^{m+1}\frac{1}{(1+q^{-1})r}\partial^{q^{-1}}_rf\big(r^2\big)+\left[\frac{m}{2}\right]_{q^2}\!\int_0^\lambda \! d_qr r^{m-1}f\big(q^{-2}r^2\big)\right]\\
=\frac{2\left(\Gamma_{q^2}(\frac{1}{2})\right)^{m}C^{ji}}{\Gamma_{q^2}(1+\frac{m}{2})(1+q)}\left[q^{m}\int_0^\lambda d_qr r^{m}\partial^q_rf\big(q^{-2}r^2\big)+\left[m\right]_q\int_0^\lambda d_qr r^{m-1}f\big(q^{-2}r^2\big)\right]\\
=\frac{2\left(\Gamma_{q^2}(\frac{1}{2})\right)^{m}C^{ji}}{\Gamma_{q^2}(1+\frac{m}{2})(1+q)}\left[\lambda^mf\big(q^{-2}\lambda^2\big)-\lim_{r\to0}r^m f\big(q^{-2}r^2\big)\right].
\end{gather*}
When we substitute $f(\bold{x}^2)=\bold{x}^{2l}E_{q^2}(-\beta \bold{x}^2)$ with $l\in\mN$ and $\beta\in\mR^+$ and use equation \eqref{zeroes} we obtain
\begin{gather*}
\int_{\mB^m_q\big(\frac{1}{\sqrt{(1-q^2)\beta}}\big)}\overline{\partial^j}\,\bold{x}^{2l}S_k(\bold{x})\,E_{q^2}\big({-}\beta \bold{x}^2\big) = 0\qquad \forall\, k,l.
\end{gather*}
This proves the second part of the theorem because of the Fischer decomposition in Lem\-ma~\ref{Fischer}.
\end{proof}

\begin{remark}
\label{gengauss}
The properties
\begin{gather*}
\int_{\gamma\cdot\mR^m_q}\overline{\partial^j}=0\qquad \mbox{on}\quad  \cP\otimes e_{q^2}\big({-}\alpha\bold{x}^2\big)\quad \mbox{for}\quad \alpha,\gamma\in\mR^+\qquad\mbox{and}\\
\int_{\mB^m_q\big(\frac{1}{\sqrt{(1-q^2)\beta}}\big)}\partial^j=0\qquad \mbox{on}\quad \cP\otimes E_{q^2}\big({-}\beta\bold{x}^2\big)\quad\mbox{for}\quad\beta\in\mR^+
\end{gather*}
also hold. They do not correspond to Gaussian-induced integration for those calculi in the strict sense. However it is a straightforward generalization, the generalized Gaussian satisf\/ies the relation
\begin{gather*}
-\eta^i_{ j}\partial^jg_{\eta}(\bold{x}) = x^i g_\eta(q\bold{x})
\end{gather*}
(or $-\eta^i_{ j}\overline{\partial^j}g_{\eta}(\bold{x})=x^i g_\eta(q^{-1}\bold{x})$) in stead of equation~\eqref{Gaussian}.
\end{remark}

Since $\int_{\gamma\cdot\mR^m_q}E_{q^2}(-\beta\bold{x}^2)$ will be inf\/inite for general $\gamma$, similar to Lemma~\ref{intexpinf}, the f\/inite integration needs to be used on $\cP\otimes E_{q^2}(-\beta\bold{x}^2)$. As in equation~\eqref{intinfeind} this integration corresponds to a specif\/ic inf\/inite integration,
\begin{gather*}
\label{intbetainf}
\int_{\mB^m_q\big(\frac{1}{\sqrt{(1-q^2)\beta}}\big)} = \int_{\big(\frac{1}{\sqrt{(1-q^2)\beta}}\big)\cdot\mR^m_q}\qquad\mbox{on}\quad \cP\otimes E_{q^2}\big({-}\beta\bold{x}^2\big)\qquad\mbox{for}\quad\beta\in\mR^+.
\end{gather*}
Theorem~\ref{2indint} therefore shows that for integration on $\cP\otimes e_{q^2}(-\alpha\bold{x}^2)$ there is a bigger choice since~$\gamma$ does not depend on $\alpha$. However, it is straightforward to calculate
\begin{gather*}
\int_{\gamma\cdot\mR^m_q}\bold{x}^{2l}S_ke_{q^2}\big({-}\alpha\bold{x}^2\big)=
\delta_{k0}\frac{\Gamma_{q^2}(\frac{m}{2}+l)}{\Gamma_{q^2}(\frac{m}{2})\alpha^lq^{\frac{1}{2}l(l+\frac{m}{2}-1)}}
\int_{\gamma\cdot\mR^m_q}e_{q^2}\big({-}\alpha\bold{x}^2\big),
\end{gather*}
which implies the integrals for dif\/ferent $\gamma$ on $\cP\otimes e_{q^2}(-\alpha\bold{x}^2)$ are proportional to each other.

In the sense of equation~\eqref{Gaussian2} the only dif\/ference between the integrations for dif\/ferent~$\gamma$ is the value $I(g_\eta)$ while the functional~$Z$ does not depend on~$\gamma$. For strict Gaussian-induced integration in the unbarred case, only one choice gives an~$I(g_\eta)$ which is f\/inite.

\subsection[Bochner's relations for the Fourier transform on $\mR^m_q$]{Bochner's relations for the Fourier transform on $\boldsymbol{\mR^m_q}$}\label{boch}

The exponential $\exp_{\hat{R}}(\bold{x}|\bold{y})$ on a quantum space satisf\/ies
\begin{gather}
\label{derFK}
\partial_x^j\exp_{\hat{R}}(\bold{x}|\bold{y}) = \exp_{\hat{R}}(\bold{x}|\bold{y})y^j,
\end{gather}
see \cite{MR1303081, MR1235979, MR2099760}. From now on $\exp_{\hat{R}}(\bold{x}|\bold{y})$ stands for the exponential on $\mR^m_q$. It is uniquely determined from equation~\eqref{derFK} and the normalization $\exp_{\hat{R}}(0|\bold{y})=1$. In order to def\/ine the Fourier transform according to~\cite{MR1303081} the Fourier kernel needs to be combined with the Gaussian induced integrations for the unbarred calculus in Theorem~\ref{2indint} and evaluated on  $\cP\otimes e_{q^2}(-\alpha\bold{x}^2)$. It turns out that the Fourier transform def\/ined by the generalized Gaussian-induced integration for the unbarred calculus in Remark~\ref{gengauss} will lead to interesting properties, see Section~\ref{behave}. This choice also corresponds to the one dimensional theory in equation~\eqref{1dFour}. The Fourier transform can be def\/ined on each space $\cP\otimes E_{q^2}(-\beta \frac{\bold{x}^2}{\mu})$. First we will extend this space. Def\/ine
\begin{gather*}
\overline{\cV}_\beta =\cP\otimes E_{q^2}\left(-\beta \frac{\bold{x}^2}{\mu}\right),\quad\beta\in\mR^+ \qquad \mbox{and} \qquad \cV_\alpha=\cP\otimes e_{q^2}\left(-\alpha\frac{q^2\bold{x}^2}{\mu}\right),\qquad\alpha\in\mR^+.
\end{gather*}
The dif\/ferent spaces $\cV_\alpha$ (or $\overline{\cV}_\beta$) are not necessarily disjunct since equation~\eqref{aflexp} implies
\begin{gather*}
\cV_{q^{-2j}\alpha}\subset\cV_{q^{-2j-2}\alpha} \qquad \mbox{and} \qquad \overline{\cV}_{q^{2j}\beta}\subset\overline{\cV}_{q^{2j+2}\beta}\qquad\mbox{for}\quad j\in\mN.
\end{gather*}
Therefore we def\/ine $\cV_{[\alpha]}=\cup_{j=0}^\infty \cV_{q^{-2j}\alpha}$ and $\overline{\cV}_{[\beta]}=\cup_{j=0}^\infty \overline{\cV}_{q^{2j}\beta}$ for $\alpha,\beta\in\mR^+$. Since, for $j\in\mN$, $\cV_{q^{2j}\alpha}\subset\cV_{\alpha}$,  $\cV_{[\alpha]}$ can also be identif\/ied with $\cup_{j=-\infty}^\infty \cV_{q^{-2j}\alpha}$ and $\overline{\cV}_{[\beta]}$ with $\cup_{j=-\infty}^\infty \overline{\cV}_{q^{2j}\beta}$.

\begin{definition}
\label{defFour1}
The f\/irst Fourier transform $\overline{\cF}^{\pm}_{\mR^m_q}$ on $\mR^m_q$ is a map $\overline{\cV}_{[\beta]}\to\cV_{[1/\beta]}$ for each $\beta\in\mR^+$,
\begin{gather*}
\overline{\cF}^\pm_{\mR^m_q}[f](\bold{y}) = \frac{1+q}{2\mu^{\frac{m}{2}}\left(\Gamma_{q^2}(\frac{1}{2})\right)^{m}}\int_{\frac{1}{\sqrt{(1-q^2)\beta}}\cdot\mR^m_q}f(\bold{x})\exp_{\hat{R}}(\pm i\bold{x}|\bold{y}).
\end{gather*}
\end{definition}
In the subsequent Corollary~\ref{BochFour}, it will be proven that the Fourier transform does indeed map elements of $\overline{\cV}_{[\beta]}$ to $\cV_{[1/\beta]}$. First we need the following technical lemma. For a polynomial~$P$, we def\/ine $[P]^j_l$ by the equation $\sum_l [P]^j_l\partial_x^l=\partial_x^j P-\left[\partial^j_xP\right]$ in ${\rm Dif\/f}(\mR^m_q)$. So in the undeformed Euclidean case, $[P]^j_l=P\delta_l^j$.
\begin{lemma}\label{Svrexp}
For $S_k\in\cS_k$ the relation $\sum_{j,l}[\partial_jS_k(\bold{x})]^j_lx^l=[k]_{q^2}S_k(\bold{x})$ holds with $\partial_j=C_{jk}\partial^k$.
\end{lemma}

\begin{proof}
The lemma is proven by calculating the expression $\sum_j [\partial_x^j[\partial_jS_k(\bold{x})]\bold{x}^2 ]$ in two dif\/ferent ways. The expression is equal to
\begin{gather*}
\sum_{j,l}\left[[\partial_jS_k(\bold{x})]^j_l\partial_x^l\bold{x}^2\right] = \sum_{j,l}[\partial_jS_k(\bold{x})]^j_l\mu x^l.
\end{gather*}
It can also be calculated using formulas \eqref{Laplx} and \eqref{Eharm}
\begin{gather*}
\sum_j\left[\partial_x^j\bold{x}^2[\partial_jS_k(\bold{x})]\right] = \sum_j\mu x^j[\partial_jS_k(\bold{x})]+\sum_jq^2\bold{x}^2\left[\partial_x^j[\partial_jS_k(\bold{x})]\right]=\mu [k]_{q^2}S_k(\bold{x}),
\end{gather*}
which proves the lemma.
\end{proof}

\begin{remark}
Using the same techniques for a general polynomial $P_k\in\cP_k$ yields
\begin{gather*}
\sum_{j,l}\left[\partial_j P_k(\bold{x})\right]^j_lx^l=[k]_{q^2}P_k(\bold{x})+\frac{(q^2-1)}{\mu^2}\bold{x}^2\Delta P_k(\bold{x}).
\end{gather*}
\end{remark}

Equations~\eqref{PizzBessel} and~\eqref{derFK} imply that the quantum sphere integral of the Fourier kernel will yield the f\/irst $q$-Bessel function. This result can be generalized by introducing spherical harmonics in the integration.

\begin{theorem}
\label{FHexp}
For $S_k\in\cS_k$, the following relation holds,
\begin{gather*}
\int_{\mS^{m-1}_q,\bold{x}}S_k(\bold{x})\exp_{\hat{R}}(i\bold{x}|\bold{y}) = 2\mu^{m/2-1} \Big(\Gamma_{q^2}\Big(\frac{1}{2}\Big)\Big)^{m}i^k\frac{J^{(1)}_{\frac{m}{2}+k-1}\big(\frac{1+q}{\mu}\sqrt{\bold{y}^2}|q^2\big)}
{\big(\sqrt{\bold{y}^2}\big)^{\frac{m}{2}+k-1}}S_k(\bold{y}).
\end{gather*}
\end{theorem}

\begin{proof}
First we prove that the relation
\begin{gather}
\label{DSexp}
\Delta_{\bold{x}}^{k+l}S_k(\bold{x})\exp_{\hat{R}}(i\bold{x}|\bold{y})
=i^{k+2l}\frac{[k+l]_{q^2}!\mu^k}{[l]_{q^2}!}\exp_{\hat{R}}(i\bold{x}|\bold{y})\bold{y}^{2l}S_k(\bold{y})+\cdots
\end{gather}
holds, where $\cdots$ stands for terms of the form $P(\bold{x})\exp_{\hat{R}}(i\bold{x}|\bold{y})q^2(\bold{y})$ with $P\in\oplus_{j>0}\cP_j$. In case $k=0$, equation~\eqref{derFK} implies that $\Delta_{\bold{x}}^{l}\exp_{\hat{R}}(i\bold{x}|\bold{y})=(-1)^l\exp_{\hat{R}}(i\bold{x}|\bold{y})\bold{y}^{2l}$ holds. Now we proceed by induction on $k$. Assuming equation \eqref{DSexp} holds for $k-1$ we calculate, using equations~\eqref{Eharm} and~\eqref{Laplx}
\begin{gather*}
\Delta_{\bold{x}}^{k+l}S_k(\bold{x})\exp_{\hat{R}}(i\bold{x}|\bold{y})
= \frac{1}{[k]_{q^2}}\Delta_{\bold{x}}^{k+l}x^j\left[\partial_{x^j}S_k(\bold{x})\right]\exp_{\hat{R}}(i\bold{x}|\bold{y})\\
\phantom{\Delta_{\bold{x}}^{k+l}S_k(\bold{x})\exp_{\hat{R}}(i\bold{x}|\bold{y})}{}
=\frac{1}{[k]_{q^2}}q^{2k+2l}x^j\Delta_{\bold{x}}^{k+l}\left[\partial_{j}S_k(\bold{x})\right]\exp_{\hat{R}}(i\bold{x}|\bold{y})\\
\phantom{\Delta_{\bold{x}}^{k+l}S_k(\bold{x})\exp_{\hat{R}}(i\bold{x}|\bold{y})=}{}
+\frac{[k+l]_{q^2}}{[k]_{q^2}}\mu\Delta_{\bold{x}}^{k+l-1}\left[\partial_{j}S_k(\bold{x})\right]^j_t\partial_{x}^t\exp_{\hat{R}}(i\bold{x}|\bold{y})\\
\phantom{\Delta_{\bold{x}}^{k+l}S_k(\bold{x})\exp_{\hat{R}}(i\bold{x}|\bold{y})}{}
=\cdots+i^{k-1+2l}\frac{[k+l]_{q^2}!}{[l]_{q^2}!}\mu^{k}\exp_{\hat{R}}(i\bold{x}|\bold{y})\bold{y}^{2l}\frac{1}{[k]_{q^2}}\left[\partial_{j}S_k(\bold{y})\right]^j_liy^l.
\end{gather*}
Lemma \ref{Svrexp} then proves that equation~\eqref{DSexp} holds for every $k$. We also used the fact that for $S_p\in\cS_p$, $[S_p]^j_l\in\cS_p$ which follows from the fact that $\partial_x^j$ and $\Delta$ commute. Equation \eqref{DSexp} then yields
\begin{gather*}
\int_{\mS^{m-1}_q,\bold{x}}S_k(\bold{x})\exp_{\hat{R}}(i\bold{x}|\bold{y})\\
 \qquad{} = \sum_{j=0}^\infty\frac{2\left(\Gamma_{q^2}(\frac{1}{2})\right)^{m}}{\mu^{2j+2k}[j+k]_{q^2}!
\Gamma_{q^2}(j+k+\frac{m}{2})}\big(\Delta^{j+k}_{\bold{x}}S_k(\bold{x})\exp_{\hat{R}}(i\bold{x}|\bold{y})\big)(\bold{x}=0)\\
 \qquad{} = i^{k}\frac{2\left(\Gamma_{q^2}(\frac{1}{2})\right)^{m}}{\mu^k}
 \sum_{j=0}^\infty\frac{(-1)^j}{[j]_{q^2}!\Gamma_{q^2}(j+k+\frac{m}{2})}\frac{\bold{y}^{2j}}{\mu^{2j}}S_k(\bold{y}).
\end{gather*}
Comparing this with equation \eqref{Bessel1} proves the theorem.
\end{proof}

This theorem allows to calculate the Fourier transform of an element of $\cP\otimes E_{q^2}(-\beta \frac{\bold{x}^2}{\mu})$ inside one of the irreducible blocks of its $\cU_q(\mathfrak{so}(m))$-decomposition.
\begin{corollary}
\label{BochFour}
The Fourier transform in Definition~{\rm \ref{defFour1}} of a function
\[
S_k(\bold{x})\psi\big(\bold{x}^2\big)\in\cP\otimes E_{q^2}\left(-\beta \frac{\bold{x}^2}{\mu}\right)\subset \overline{\cV}_{[\beta]}
\]
 with $S_k\in\cS_k$ is given by
\begin{gather*}
\overline{\cF}^\pm\big[S_k(\bold{x})\psi\big(\bold{x}^2\big)\big](\bold{y}) = (\pm i)^kS_k(\bold{y})\overline{\cF}^{q,\beta}_{\frac{m}{2}+k-1}[\psi]\big(\bold{y}^2\big)\\
\qquad{}  = (\pm i)^k\frac{1+q}{\mu}\left[\int_{0}^{\sqrt{\frac{\mu}{(1-q^2)\beta}}}d_qr \,r^{m+2k-1}\psi\big(r^2\big)\frac{J^{(1)}_{\frac{m}{2}+k-1}\big(\frac{1+q}{\mu}rt|q^2\big)}{\left(rt\right)^{\frac{m}{2}+k-1}}
\right]_{t^2=\bold{y}^2}S_k(\bold{y})
\end{gather*}
with $r$ and $t$ two real commuting variables.
\end{corollary}

These formulae are the Bochner's relations for the Fourier transform on $\mR^m_q$. For the classical Bochner's relations, see e.g.~\cite{MR1151617}.

This corresponds exactly to the one dimensional case for $\beta=1$. Since $\lim\limits_{m\to1}\mu=1+q$, Corollary~\ref{BochFour} for $m=1$ and $k=0$ reduces to
\begin{gather*}
\frac{\sqrt{1+q}}{2\Gamma_{q^2}(\frac{1}{2})}\int_{-\frac{1}{\sqrt{1-q}}}^{\frac{1}{\sqrt{1-q}}}d_qx\,\psi\big(x^2\big)e_q(ixt) = \left[\int_{0}^{\frac{1}{\sqrt{1-q}}}d_qr \psi(r^2)\frac{J^{(1)}_{-\frac{1}{2}}(rt|q^2)}{(rt)^{-\frac{1}{2}}}\right] \quad\mbox{for}\quad t^2<\frac{1}{1-q}.
\end{gather*}
Corollary \ref{BochFour} for $m=1$, $k=1$ and $S_1=x$ is
\begin{gather*}
\frac{\sqrt{1+q}}{2\Gamma_{q^2}(\frac{1}{2})}\int_{-\frac{1}{\sqrt{1-q}}}^{\frac{1}{\sqrt{1-q}}}d_qx\,x\,\psi\big(x^2\big)e_q(ixt) = i\left[ \! \int_{0}^{\frac{1}{\sqrt{1-q}}} \!\!d_qrr^2 \psi(r^2)\frac{J^{(1)}_{\frac{1}{2}} (rt|q^2 )}{(rt)^{\frac{1}{2}}}\right]t
\quad \mbox{for}\quad t^2<\frac{1}{1-q}.
\end{gather*}
This agrees with
\[
e_q(iu)=\Gamma_{q^2}\Big(\frac{1}{2}\Big)\left(\frac{u}{1+q}\right)^{(1/2)}\left[J_{-\frac{1}{2}}^{(1)}\big(u|q^2\big)
+iJ_{\frac{1}{2}}^{(1)}\big(u|q^2\big)\right],
 \]
 which can be easily calculated.

\section{Properties of the Fourier transform}
\label{behave}

The Fourier transform is determined by its Bochner's relations, see Corollary \ref{BochFour}. The second Fourier transform is immediately def\/ined here by its Bochner's relations.
\begin{definition}
\label{defFour}
The second Fourier transform $\cF^{\pm}_{\mR^m_q}$ on $\mR^m_q$ is a map $\cV_{[\alpha]}\to\overline{\cV}_{[1/\alpha]}$ for each $\alpha\in\mR^+$. For a function $S_k(\bold{y})\psi(\bold{y}^2)\in\cP\otimes E_{q^2}\big({-}\frac{\bold{y}^2}{\mu\alpha}\big)$ the transform is given by
\begin{gather*}
\cF^{\pm}_{\mR^m_q}\big[S_k(\bold{y})\psi(\bold{y}^2)\big](\bold{x}) = (\pm i)^kS_k(\bold{x})\frac{1}{c(\sqrt{\alpha}\gamma)}\cF^{q,\gamma}_{\frac{m}{2}+k-1}\left[\psi\right]\big(\bold{x}^2\big),
\end{gather*}
for an arbitrary $\gamma\in\mR^+$ with
\begin{gather*}
c\big(\sqrt{\alpha}\gamma\big) = \frac{q^{(\frac{m}{2}-1)\frac{m}{2}}}{\Gamma_{q^2}(\frac{m}{2})}\int_0^{\frac{\alpha q^2\gamma^2 }{\mu}\cdot\infty}d_{q^2}u u^{\frac{m}{2}-1}E_{q^2}(-u)=d\left(\gamma\sqrt{\frac{\alpha}{\mu}},\frac{m}{2}-1\right)\\
 \phantom{c\big(\sqrt{\alpha}\gamma\big)}{}
 = d\left(\gamma\sqrt{\frac{q^{-2j}\alpha}{\mu}},\frac{m}{2}+k-1\right)\qquad\mbox{for} \quad j,k\in\mN.
\end{gather*}
\end{definition}
The second Fourier transform does not depend on the choice of $\gamma$ as can be seen from the expressions in the subsequent Theorem~\ref{Fourinv}. From the properties of $c(\sqrt{\alpha}\gamma)$ it is clear that the def\/inition does not depend on which $\cV_\alpha$ the element of $\cV_{[\alpha]}$ is chosen to be in. Although the Fourier transform on each space~$\cV_{[\alpha]}$ is denoted by the same symbol, each Fourier transform should be regarded as an independent operator. In Section~\ref{FourHO} these dif\/ferent transforms will be combined in order to construct the Fourier transform on the Hilbert space corresponding to the harmonic oscillator.

\begin{theorem}
\label{Fourinv}
The Fourier transforms in Definitions~{\rm \ref{defFour1}} and~{\rm \ref{defFour}} are each others inverse, i.e.\ for each $\alpha,\beta\in\mR^+$
\begin{gather*}
\overline{\cF}_{\mR^m_q}^{\mp}\circ\cF^{\pm}_{\mR^m_q} = id_{\cV_{[\alpha]}}\qquad\mbox{and}\qquad
\cF_{\mR^m_q}^{\mp}\circ\overline{\cF}^{\pm}_{\mR^m_q} = id_{\overline{\cV}_{[\beta]}}.
\end{gather*}
\end{theorem}

\begin{proof}
This is a consequence of Corollary~\ref{BochFour} and Theorem~\ref{Hankelinv}. It can also be obtained directly from the relations,
\begin{gather*}
\overline{\cF}^{\pm}_{\mR^m_q}\left[\cL_j^{(\frac{m}{2}+k-1)}\left(\frac{\bold{x}^2}{\alpha\mu}|q^2\right)
S_k(\bold{x})E_{q^2}\left(-\frac{\bold{x}^2}{\alpha\mu}\right)\right](\bold{y})\\
 \qquad{} = (\pm i)^k\alpha^{\frac{m}{2}+k+j}C_j\bold{y}^{2j}S_k(\bold{y})e_{q^2}\left(-\alpha\frac{q^2\bold{y}^2}{\mu}\right),\\
\cF^{\mp}_{\mR^m_q}\left[\alpha^{\frac{m}{2}+k+j}C_j\bold{y}^{2j}S_k(\bold{y})e_{q^2}\left(-\alpha\frac{q^2\bold{y}^2}{\mu}\right)\right](\bold{x})\\
 \qquad{} = (\mp i)^k\cL_j^{(\frac{m}{2}+k-1)}\left(\frac{\bold{x}^2}{\alpha\mu}|q^2\right)S_k(\bold{x})E_{q^2}\left(-\frac{\bold{x}^2}{\alpha\mu}\right),
\end{gather*}
with $C_j=\frac{q^{2(j+1)(j+\frac{m}{2}+k)}}{[j]_{q^2}!\mu^j}$. These follow immediately from the calculations before Theorem~\ref{Hankelinv}.
\end{proof}

Partial derivatives and multiplication with variables are operations which are def\/ined on~$\cV_\alpha$ and $\cV_{\beta}$ and therefore also on~$\cV_{[\alpha]}$ and~$\cV_{[\beta]}$. In this section we investigate how they interact with the Fourier transforms.

\begin{theorem}
\label{Fourierx}
For $f\in\overline{\cV}_{[\beta]}$ and $g\in\cV_{[\alpha]}$ with $\alpha,\beta\in\mR^+$, the following relations hold:
\begin{alignat*}{3}
&(i) && \overline{\cF}_{\mR_q^m}^{\pm}[x^jf(\bold{x})](\bold{y}) = \mp i\partial_y^j\overline{\cF}_{\mR_q^m}^{\pm}[f(\bold{x})](\bold{y}),& \\
&(ii) & &\cF^{\pm}_{\mR_q^m}[y^jg(\bold{y})](\bold{x}) = \frac{\mp i}{q^m}\overline{\partial_x^j}\cF^{\pm}_{\mR_q^m}[g(\bold{y})](\bold{x}),& \\
&(iii)\quad &&\overline{\cF}^{\pm}_{\mR^m_q}\big[\overline{\partial^j_x}f(\bold{x})\big](\bold{y}) = \mp iq^my^j\overline{\cF}^{\pm}_{\mR^m_q}\left[f(\bold{x})\right](\bold{y}),& \\
&(iv)&& \cF^{\pm}_{\mR^m_q}\left[\partial^j_yg(\bold{y})\right](\bold{x}) = \mp ix^j\cF^{\pm}_{\mR^m_q}\left[g(\bold{y})\right](\bold{x}).&
\end{alignat*}
\end{theorem}

\begin{proof}
In order to prove $(i)$ we choose $f(\bold{x})$ of the form $S_k(\bold{x})\psi(\bold{x}^2)\in \cP\otimes E_{q^2}\big({-}\beta\frac{\bold{x}^2}{\mu}\big)$ for an arbitrary $\beta\in\mR^+$. We use the Fischer decomposition of $x^jS_k(\bold{x})$ in Lemma~\ref{xS} and def\/ine $\psi'(r^2)=r^2\psi(r^2)$. Then,
\begin{gather*}
\overline{\cF}_{\mR_q^m}^{\pm}\big[x^jS_k(\bold{x})\psi\big(\bold{x}^2\big)\big](\bold{y}) = (\pm i)^{k+1}y^jS_{k}(\bold{y})\overline{\cF}^{q,\beta}_{\frac{m}{2}+k}[\psi]\big(\bold{y}^2\big)\\
\qquad{}
+(\pm i)^{k-1}\frac{\bold{y}^2\partial^jS_{k}(\bold{y})}{\mu[\frac{m}{2}+k-1]_{q^2}}\overline{\cF}^{q,\beta}_{\frac{m}{2}+k}[\psi]\big(\bold{y}^2\big)
 + (\pm i)^{k-1}\frac{\partial^jS_{k}(\bold{y})}{\mu[\frac{m}{2}+k-1]_{q^2}}\overline{\cF}^{q,\beta}_{\frac{m}{2}+k-2}[\psi']\big(\bold{y}^2\big)
\end{gather*}
holds. Applying Lemma \ref{eigHankF}$(i)$ then yields
\begin{gather*}
\overline{\cF}_{\mR_q^m}^{\pm}\big[x^jS_k(\bold{x})\psi\big(\bold{x}^2\big)\big](\bold{y}) \\
\qquad{} = (\pm i)^{k+1}y^jS_{k}(\bold{y})\overline{\cF}^{q,\beta}_{\frac{m}{2}+k}[\psi]\big(\bold{y}^2\big)+(\pm i)^{k-1}\left[\partial^jS_{k}(\bold{y})\right]\overline{\cF}^{q,\beta}_{\frac{m}{2}+k-1}[\psi]\big(q^2\bold{y}^2\big).
\end{gather*}
Now we calculate the right-hand side, using equations \eqref{partialr} and \eqref{berpx2} and Lemma~\ref{eigHankF}$(ii)$,
\begin{gather*}
\mp i\partial_y^j\overline{\cF}_{\mR_q^m}^{\pm}\big[S_k(\bold{x})\psi\big(\bold{x}^2\big)\big](\bold{y})\\
 \qquad {} = -(\pm i)^{k+1}y^jS_k(\bold{y})\mu\partial_{\bold{y}^2}^{q^2}\overline{\cF}^{q,\beta}_{\frac{m}{2}+k-1}\left[\psi\right]\big(\bold{y}^2\big)
 + (\pm i)^{k-1}\left[\partial^jS_k(\bold{y})\right]\overline{\cF}^{q,\beta}_{\frac{m}{2}+k-1}\left[\psi\right]\big(q^2\bold{y}^2\big)\\
\qquad{} = (\pm i)^{k+1}y^jS_k(\bold{y})\overline{\cF}^{q,\beta}_{\frac{m}{2}+k}\left[\psi\right]\big(\bold{y}^2\big)
 + (\pm i)^{k-1}\left[\partial^jS_k(\bold{y})\right]\overline{\cF}^{q,\beta}_{\frac{m}{2}+k-1}\left[\psi\right]\big(q^2\bold{y}^2\big).
\end{gather*}
This proves property $(i)$.

Property $(ii)$ can be calculated using the exact same techniques. Consider $S_k(\bold{y})\psi(\bold{y}^2)\in\cP\otimes e_{q^2}\big({-}\alpha\frac{q^2\bold{y}^2}{\mu}\big)$. Combining Lemma~\ref{xS} for the barred calculus with Lemma~\ref{eigHankF}$(i)$ yields
\begin{gather*}
c\big(\sqrt{\alpha}\gamma\big)\cF_{\mR_q^m}^{\pm }\big[y^jS_k(\bold{y})\psi\big(\bold{y}^2\big)\big](\bold{x})\\
 \qquad{} = (\pm i)^{k+1}x^jS_{k}(\bold{x})\cF^{q,\gamma}_{\frac{m}{2}+k}[\psi]\big(\bold{x}^2\big)
 + \frac{(\pm i)^{k-1}}{q^{m}}\big[\overline{\partial^j}S_{k}(\bold{x})\big]\cF^{q,\gamma}_{\frac{m}{2}+k-1}[\psi]\big(q^{-2}\bold{x}^2\big).
\end{gather*}
Equation \eqref{partialrbar} and Lemma~\ref{eigHankF}$(ii)$ imply that this expression is equal to
\[
c(\sqrt{\alpha}\gamma)\frac{\mp i}{q^m}\overline{\partial^j}\cF^{\pm}_{\mR_q^m}[f(\bold{y})](\bold{x}).
\]

Properties $(iii)$ and $(iv)$ follow from properties $(i)$ and $(ii)$ and Theorem \ref{Fourinv}. As an illustration we calculate property $(iv)$ directly. Consider $S_k(\bold{y})\psi(\bold{y}^2)\in\cP\otimes e_{q^2}\big({-}\alpha\frac{q^2\bold{y}^2}{\mu}\big)$ for an arbitrary $\alpha\in\mR^+$. Using equation \eqref{partialr} in the left-hand side of the second property yields
\begin{gather*}
\cF^{\pm}_{\mR^m_q}\big[\partial^j_yS_k(\bold{y})\psi\big(\bold{y}^2\big)\big](\bold{x}) =  \cF^{\pm}_{\mR^m_q}\big[\mu y^j\big(\partial^{q^2}_{\bold{y}^2}\psi\big(\bold{y}^2\big)\big)S_k(\bold{y})\big](\bold{x})+ \cF^{\pm}_{\mR^m_q}\big[(\partial^jS_k)(\bold{y})\psi\big(q^2\bold{y}^2\big)\big](\bold{x}).
\end{gather*}
The property $\int_0^{\gamma\cdot\infty}d_qt f(qt)=\frac{1}{q}\int_0^{\gamma\cdot\infty}d_qt f(t)$ yields
\begin{gather*}
c\big(\sqrt{\alpha}\gamma\big)\cF^{\pm}_{\mR^m_q}\big[(\partial^jS_k)(\bold{y})\psi\big(q^2\bold{y}^2\big)\big](\bold{x}) = \frac{(\pm i)^{k-1}}{q^{m+2k-2}}(\partial^jS_k)(\bold{x})\cF^{q,\gamma}_{\frac{m}{2}+k-2}\left[\psi\right]\big(q^{-2}\bold{x}^2\big).
\end{gather*}
For $S_{k+1}\in\cS_{k+1}$, Lemma \ref{eigHankF}$(iii)$ implies
\begin{gather*}
c(\sqrt{\alpha}\gamma)\mu\cF^{\pm,\gamma}_{\mR^m_q}\big[ S_{k+1}(\bold{y})\partial^{q^2}_{\bold{y}^2}\psi\big(\bold{y}^2\big)\big](\bold{x}) = (\pm i)^{k-1}S_{k+1}(\bold{y})\cF^{q,\gamma}_{\frac{m}{2}+k-1}[\psi]\big(\bold{x}^2\big)
\end{gather*}
and for $S_{k-1}\in\cS_{k-1}$ Lemma \ref{lemHankelextra} yields
\begin{gather*}
c\big(\sqrt{\alpha}\gamma\big)\mu \cF^{\pm}_{\mR^m_q}\big[ S_{k-1}(\bold{y})\bold{y}^2\big(\partial^{q^2}_{\bold{y}^2}\psi\big(\bold{y}^2\big)\big)\big](\bold{x}) \\
= (\pm i)^{k-1}\bold{x}^2S_{k-1}(\bold{x})\cF^{q,\gamma}_{\frac{m}{2}+k-1}[\psi]\big(\bold{x}^2\big)
 - \frac{[\frac{m}{2}+k-1]_{q^2}}{q^{{m}+2k-2}}\mu(\pm i)^{k-1}S_{k-1}(\bold{x})\cF^{q,\gamma}_{\frac{m}{2}+k-2}[\psi]\big(q^{-2}\bold{x}^2\big).
\end{gather*}
From these calculations combined with Lemma~\ref{xS} and taking $S_{k+1}$ and $S_{k-1}$ as def\/ined by $x^jS_k=S_{k+1}+\bold{x}^2S_{k-1}$, we obtain
\begin{gather*}
c\big(\sqrt{\alpha}\gamma\big)\cF^{\pm}_{\mR^m_q}\big[\partial^j_yS_k(\bold{y})\psi\big(\bold{y}^2\big)\big](\bold{x})\\
\qquad{} = (\pm i)^{k-1}\big[S_{k+1}(\bold{y})\cF^{q,\gamma}_{\frac{m}{2}+k-1}[\psi]\big(\bold{x}^2\big)
+\bold{x}^2S_{k-1}(\bold{x})\cF^{q,\gamma}_{\frac{m}{2}+k-1}[\psi]\big(\bold{x}^2\big)\big]\\
\qquad{} = (\pm i)^{k-1}x^jS_{k}(\bold{x})\cF_{\frac{m}{2}+k-1}^{q,\gamma}[\psi]\big(\bold{x}^2\big).
\end{gather*}
This proves property $(iv)$.
\end{proof}

The behavior of the Fourier transforms with respect to the Laplacian and norm squared can be calculated from Theorem~\ref{Fourierx}. For $\overline{\cF}^{\pm}_{\mR^m_q}$ acting on each space $\overline{\cV}_{[\beta]}$ and for ${\cF}^{\pm}_{\mR^m_q}$ acting on each space ${\cV}_{[\alpha]}$, the relations
\begin{gather*}
\overline{\cF}_{\mR^m_q}^{\pm}\circ \bold{x}^2=-\Delta_{\bold{y}}\circ\overline{\cF}_{\mR^m_q}^{\pm} ,\qquad \overline{\cF}_{\mR^m_q}^{\pm}\circ {\Delta}^\ast_{\bold{x}}=-\bold{y}^2\circ\overline{\cF}_{\mR^m_q}^{\pm},\\
\cF_{\mR^m_q}^{\pm}\circ \bold{y}^2=-{\Delta}^\ast_{\bold{x}}\circ\cF_{\mR^m_q}^{\pm}, \qquad 
\cF_{\mR^m_q}^{\pm}\circ {\Delta_{\bold{y}}}=-\bold{x}^2\circ\cF_{\mR^m_q}^{\pm}
\end{gather*}
hold. This implies that the Fourier transforms map the two Hamiltonians for the harmonic oscillator in~\eqref{HOs} into each other,
\begin{gather}
\label{fourham}
\overline{\cF}_{\mR^m_q}^{\pm}\circ {h}^\ast_{\bold{x}}=h_{\bold{y}}\circ\overline{\cF}_{\mR^m_q}^{\pm} \qquad \mbox{and} \qquad \cF_{\mR^m_q}^{\pm}\circ h_{\bold{y}}={h}^\ast_{\bold{x}}\circ\cF_{\mR^m_q}^{\pm}.
\end{gather}

\section[Funk-Hecke theorem on $\mR^m_q$]{Funk--Hecke theorem on $\boldsymbol{\mR^m_q}$}

The polynomials on the quantum sphere correspond to $\cP/(R^2-1)$ with $(R^2-1)$ the ideal generated by the relation $R^2-1$. The Fischer decomposition in Lemma~\ref{Fischer} implies that this space is isomorphic to $\cS=\bigoplus_{k=0}^\infty\cS_k$. The inner product on the quantum sphere $\langle \cdot|\cdot\rangle:\cS\times\cS\to\mC$
\begin{gather}
\label{inSymm}
\langle f|g\rangle=\int_{\mS^{m-1}_q}f(g)^\ast
\end{gather}
is positive def\/inite, symmetric and $O_q(m)$-invariant, see~\cite[Proposition~14]{MR1328733}. In particular $\cS_k\perp \cS_l$ when $k\not=l$. The symmetry can be obtained from the results in~\cite{MR1239953} or from the subsequent Lemma~\ref{Sast}.

In~\cite{MR1317458} the polynomials of degree $l$ in $\bold{x}$ and $\bold{y}$, $(\bold{x};\bold{y})^{(l)}$ which satisfy
\[
\partial_x^j(\bold{x};\bold{y})^{(l)}=[l]_{q^{-2}}(\bold{x};\bold{y})^{(l-1)}y^j
 \]
 were determined. In particular this implies that the exponential on $\mR^m_q$ takes the form
\begin{gather*}
\exp_{\hat{R}}(\bold{x}|\bold{y}) = \sum_{l=0}^\infty \frac{(\bold{x};\bold{y})^{(l)}}{[l]_{q^{-2}}!}.
\end{gather*}
For later convenience we def\/ine $\langle \bold{x}|\bold{y}\rangle^{(l)}=\frac{[l]_q!}{[l]_{q^{-2}}!}(\bold{x};\bold{y})^{(l)}$, so
\begin{gather*}
\partial_x^j\langle \bold{x}|\bold{y}\rangle^{(l)} = [l]_{q}\langle \bold{x}|\bold{y}\rangle^{(l-1)}y^j.
\end{gather*}
These polynomials satisfy the following Funk--Hecke theorem, see \cite{MR0499342} for the classical version.
\begin{theorem}
\label{FHthm}
For $S_k\in\cS_k$, the relation
\begin{gather*}
\int_{\mS^{m-1}_q,\bold{x}}S_k(\bold{x})\langle \bold{x}|\bold{y}\rangle^{(l)} = \alpha_{k,l}S_k(\bold{y}) \bold{y}^{l-k}
\end{gather*}
holds with
\begin{gather*}
\alpha_{k,l} = \begin{cases}
\dfrac{2\Gamma_{q^2}(\frac{1}{2})^m[l]_q!}{\mu^l\left[\frac{l-k}{2}\right]_{q^2}!\Gamma_{q^2}\left(\frac{k+l+m}{2}\right)}\quad & \mbox{if} \ k+l \ \mbox{is  even and} \  l\ge k,\\
 0 \quad & \mbox{if} \ k+l \ \mbox{is odd},\\
 0\quad & \mbox{if} \ l<k.
 \end{cases}
\end{gather*}
\end{theorem}
\begin{proof}
This follows immediately from Theorem \ref{FHexp}.
\end{proof}

This will lead to the reproducing kernel for the spherical harmonics. First the following technical lemma is needed.
\begin{lemma}
\label{lemtechgamma}
For $l>0$ and $\alpha\in\mR$, the following relation holds:
\begin{gather*}
\sum_{j=0}^l(-1)^jq^{j(j-1)}\binom{l}{j}_{q^2}\frac{\Gamma_{q^2}(\alpha+l-j)}{\Gamma_{q^2}(\alpha+1-j)} = 0.
\end{gather*}
\end{lemma}
\begin{proof}
The calculation $\binom{l}{j}_{q^2}=\binom{l-1}{j-1}_{q^2}+q^{2j}\binom{l-1}{j}_{q^2}$ is straightforward. Applying this yields
\begin{gather*}
 C[l]:=\sum_{j=0}^l(-1)^j\binom{l}{j}_{q^2}\frac{\Gamma_{q^2}(\alpha+l-j)}{\Gamma_{q^2}(\alpha+1-j)}  = \sum_{j=0}^{l-1}(-1)^jq^{j(j-1)}\binom{l-1}{j}_{q^2}q^{2j}\frac{\Gamma_{q^2}(\alpha+l-j)}{\Gamma_{q^2}(\alpha+1-j)}\\
\phantom{C[l]=}{}
-\sum_{j=0}^{l-1}(-1)^jq^{j(j+1)}\binom{l-1}{j}_{q^2}\frac{\Gamma_{q^2}(\alpha+l-j-1)}{\Gamma_{q^2}(\alpha-j)}\\
\phantom{C[l]}{}
 = \sum_{j=0}^{l-1}(-1)^jq^{j(j-1)}\binom{l-1}{j}_{q^2}q^{2j}\frac{\Gamma_{q^2}(\alpha+l-1-j)}{\Gamma_{q^2}(\alpha+1-j)}
 \left([\alpha+l-1-j]_{q^2}-[\alpha-j]_{q^2}\right)\\
\phantom{C[l]}{}
= q^{2\alpha}[l-1]_{q^2}C[l-1].
\end{gather*}
This relation shows that $C[1]=0$ and by induction $C[l]=0$, for $l>0$.
\end{proof}

We def\/ine the coef\/f\/icients $c_j^{n,\lambda}$ of the $q$-Gegenbauer polynomials \eqref{qGeg} as
\begin{gather}
\label{coeffGeg}
C^\lambda_n\left(q;\frac{\mu}{1+q}t\right) = \sum_{j=0}^{\lfloor \frac{n}{2}\rfloor}c^{n,\lambda}_jt^{n-2j}.
\end{gather}

\begin{theorem}
The polynomials
\begin{gather*}
F_n(\bold{x}|\bold{y}) = C_n\sum_{j=0}^{\lfloor \frac{n}{2}\rfloor}c^{n,\frac{m}{2}-1}_j\bold{x}^{2j}\langle\bold{x}|\bold{y}\rangle^{(n-2j)}\bold{y}^{2j}
\end{gather*}
with $c^{n,\lambda}_j$ as defined in equation~\eqref{coeffGeg} and $C_n=\frac{[\frac{m}{2}+n-1]_{q^2}\Gamma_{q^2}(\frac{m}{2}-1)}{2\Gamma_{q^2}(\frac{1}{2})^m}$ satisfy
\begin{gather}
\label{repKern}
\int_{\mS^{m-1}_q}S_k(\bold{x})F_n(\bold{x}|\bold{y}) = \delta_{kn}S_k(\bold{y})\qquad \mbox{for} \quad S_k\in\cS_k.
\end{gather}
\end{theorem}

\begin{proof}
If $n-k$ is odd or $n<k$ then the left-hand side of equation \eqref{repKern} is zero because of the expression for $\alpha_{k,l}$ in Theorem~\ref{FHthm}. So we consider the case $n\ge k$ and $n-k$ even. The left-hand side of equation \eqref{repKern} can then be calculated using Theorem~\ref{FHthm}
\begin{gather*}
\int_{\mS^{m-1}_q}S_k(\bold{x})F_n(\bold{x}|\bold{y}) = C_n\sum_{j=0}^{\lfloor \frac{n}{2}\rfloor}c^{n,\frac{m}{2}-1}_j\int_{\mS^{m-1}_q}S_k(\bold{x})\langle\bold{x}|\bold{y}\rangle^{(n-2j)}\bold{y}^{2j}\\
\hphantom{\int_{\mS^{m-1}_q}S_k(\bold{x})F_n(\bold{x}|\bold{y})}{}
= C_n\left(\sum_{j=0}^{\frac{n-k}{2}}c^{n,\frac{m}{2}-1}_j\alpha_{k,n-2j}\right)S_k(\bold{y})\bold{y}^{n-k}.
\end{gather*}
Therefore we calculate
\begin{gather*}
\sum_{j=0}^{\frac{n-k}{2}}c^{n,\frac{m}{2}-1}_j\alpha_{k,n-2j} \\
\qquad{}= \sum_{j=0}^{\frac{n-k}{2}}\frac{(-1)^jq^{j(j-1)}}{[j]_{q^2}![n-2j]_q!}
\frac{\Gamma_{q^2}(\frac{m}{2}-1+n-j)}{\Gamma_{q^2}(\frac{m}{2}-1)}\mu^{n-2j}
\frac{2\Gamma_{q^2}(\frac{1}{2})^m[n-2j]_q!}{\mu^{n-2j}\big[\frac{n-k}{2}-j\big]_{q^2}!\Gamma_{q^2}\big(\frac{k+n-2j+m}{2}\big)}\\
 \qquad{} = \frac{2\Gamma_{q^2}(\frac{1}{2})^m}{\Gamma_{q^2}(\frac{m}{2}-1)\big[\frac{n-k}{2}\big]_{q^2}!}
 \sum_{j=0}^{\frac{n-k}{2}}(-1)^jq^{j(j-1)}\binom{\frac{n-k}{2}}{j}_{q^2}\frac{\Gamma_{q^2}(\frac{m}{2}-1+n-j)}{\Gamma_{q^2}\big(\frac{k+n+m}{2}-j\big)}.
\end{gather*}
When $n-k>0$ this expression is zero because of Lemma~\ref{lemtechgamma} for $l=(n-k)/2$ and $\alpha=(k+n+m)/2-1$. This implies
\begin{gather*}
\sum_{j=0}^{\frac{n-k}{2}}c^{n,\frac{m}{2}-1}_j\alpha_{k,n-2j} = \delta_{kn}\frac{2\Gamma_{q^2}(\frac{1}{2})^m}{\Gamma_{q^2}(\frac{m}{2}-1)}\frac{1}{\big[\frac{m}{2}+k-1\big]_{q^2}}=\frac{\delta_{kn}}{C_n},
\end{gather*}
which proves the theorem.
\end{proof}

This theorem implies that for bases $\{S_k^{(l)}\}$ of $\cS_k$, which are orthonormal with respect to the inner product in equation~\eqref{inSymm}, the reproducing kernel satisf\/ies
\begin{gather*}
F_k(\bold{x}|\bold{y}) = \sum_{l=1}^{\dim \cS_k}\big(S_k^{(l)}(\bold{x})\big)^\ast S_k^{(l)}(\bold{y}).
\end{gather*}

The reproducing kernel can be written symbolically as a $q$-Gegenbauer polynomial, keeping in mind that $\langle \bold{x}|\bold{y}\rangle^j$ should be replaced by $\langle \bold{x}|\bold{y}\rangle^{(j)}$. In~\cite{CDBS2} an overview of Gegenbauer polynomials appearing as reproducing kernels for classical, Dunkl and super harmonic analysis is given. For completeness we prove that $\big(S_k^{(l)}(\bold{x})\big)^\ast$ is still a spherical harmonic.

\begin{lemma}
\label{Sast}
The antilinear involutive antihomomorphism $\ast$ on $\cP$ satisfies $\left(\bold{x}^{2l}\cS_k\right)^\ast\subset \bold{x}^{2l}\cS_k$.
\end{lemma}

\begin{proof}
It is immediately clear that for $S_k\in\cS_k$, $\left(\bold{x}^{2l}S_k(\bold{x})\right)^\ast=\bold{x}^{2l}\left(S_k(\bold{x})\right)^\ast$ holds. Induction and equation~\eqref{Eharm} yield
\begin{gather*}
\left(S_k(\bold{x})\right)^\ast =  \frac{1}{[k]_{q^2}!}\sum_{i_1,\dots,i_k=1}^{m}x^{i_1}x^{i_2}\cdots x^{i_k}\partial^{i_1}_x\partial^{i_2}_x\cdots\partial^{i_k}_xS_k(\bold{x}).
\end{gather*}
First we will prove the relation
\begin{gather}
\label{tssvgl}
\sum_{i_l,i_{l+1},i_{l+2},\dots,i_k=1}^m\partial^{i_l}_x x^{i_{l+1}}x^{i_{l+2}}\cdots x^{i_k}\partial^{i_1}_x\partial^{i_2}_x
\cdots\partial^{i_k}_xS_k(\bold{x})=0
\end{gather}
for $i_1,\dots, i_{l-1}\in\{1,\dots,m\}$, $l=k-1,k-2,\dots,1$ by induction. This clearly holds for $l=k-1$, since equation \eqref{defpart} implies
\begin{gather*}
\sum_{i_{k-1},i_{k}=1}^m\partial^{i_{k-1}}_x x^{i_k}\partial^{i_1}_x\partial^{i_2}_x\cdots\partial^{i_k}_xS_k(\bold{x}) = \partial^{i_1}_x\partial^{i_2}_x\cdots\partial^{i_{k-2}}_x\Delta S_k(\bold{x}).
\end{gather*}
Then we assume it holds for $l+1$ and calculate
\begin{gather*}
 \sum_{i_{l},i_{l+1},\dots,i_k=1}^m\partial^{i_l}_x x^{i_{l+1}}x^{i_{l+2}}\cdots x^{i_k}\partial^{i_1}_x\partial^{i_2}_x\cdots\partial^{i_k}_xS_k(\bold{x})\\
\qquad = \sum_{i_{l},i_{l+1},\dots,i_k=1}^mC_{i_{l}i_{l+1}}x^{i_{l+2}}\cdots x^{i_k}\partial^{i_1}_x\partial^{i_2}_x\cdots\partial^{i_k}_xS_k(\bold{x})\\
\qquad\quad{} + q\sum_{i_{l},i_{l+1},\dots,i_k=1}^m\sum_{s,t=1}^m(\hat{R}^{-1})^{i_l,i_{l+1}}_{s,t}x^s\partial^{t}_x x^{i_{l+2}}\cdots x^{i_k}\partial^{i_1}_x\partial^{i_2}_x\cdots\partial^{i_k}_xS_k(\bold{x}).
\end{gather*}
The f\/irst line after the equality is zero since $S_k\in\cS_k$. The second one can be simplif\/ied by using relation~\eqref{Rxx}:
\begin{gather*}
\sum_{s=1}^mx^s\sum_{t,i_{l+2},\dots,i_k=1}^m\partial^{t}_x x^{i_{l+2}}\cdots x^{i_k}\partial^{i_1}_x\partial^{i_2}_x\cdots\partial^{i_{l-1}}_x\partial^{s}_x\partial^{t}_x\partial^{i_{l+2}}_x\cdots\partial^{i_{k}}_xS_k(\bold{x}).
\end{gather*}
This is zero because of the induction step.

Then we use equation \eqref{Laplx} to calculate
\begin{gather*}
\Delta S_k^\ast(\bold{x}) = \mu\frac{1}{[k]_{q^2}!}\sum_{i_1,\dots,i_k=1}^{m}\partial^{i_1}x^{i_2}\cdots x^{i_k}\partial^{i_1}_x\partial^{i_2}_x\cdots\partial^{i_k}_xS_k(\bold{x})\\
\phantom{\Delta S_k^\ast(\bold{x}) =}{} + q^2\frac{1}{[k]_{q^2}!}\sum_{i_1,\dots,i_k=1}^{m}x^{i_1}\Delta x^{i_2}\cdots x^{i_k}\partial^{i_1}_x\partial^{i_2}_x\cdots\partial^{i_k}_xS_k(\bold{x})
\end{gather*}
which implies $S_k^\ast\in\cS_k$ by using equation \eqref{tssvgl} consecutively.
\end{proof}

\section{The Fourier transform on the Hilbert space\\ of the harmonic oscillator}\label{FourHO}

The Fourier transforms have been def\/ined on specif\/ic spaces of polynomials weighted with Gaussians. The Fourier transform can be extended to the Hilbert space structure developed in~\cite{MR1239953}. First we repeat the basic ideas of~\cite{MR1288667, MR1239953}. Def\/ine the functions
\begin{gather*}
\psi_0=e_{q^2}\left(-\frac{\bold{x}^2}{q^{m/2}\mu}\right)\qquad \mbox{and} \qquad \overline{\psi}_0=E_{q^2}\left(-\frac{q^{m/2}q^2\bold{x}^2}{\mu}\right),
\end{gather*}
which are the ground states corresponding to the Hamiltonians in equation \eqref{HOs}. A $q$-deformation of the raising operators is given by
\begin{gather*}
a^{j+}_n=b_n(q)\big(x^j-q^{2-n-\frac{m}{2}}\partial^j\big)\Lambda^{-\frac{1}{4}}\qquad
\mbox{and} \qquad \overline{a}^{j+}_n=\overline{b}_n(q)\big(x^j-q^{n-2-\frac{m}{2}}\overline{\partial^j}\big)\Lambda^{\frac{1}{4}},
\end{gather*}
with (for now) undetermined coef\/f\/icients $b_n(q)$ and $\overline{b}_n(q)$. In~\cite{MR1239953} the relation $b_n(q)=\overline{b}_n(q^{-1})$ was assumed, which we do not assume here.
These operators can be used to construct the functions
\begin{gather*}
\psi^{i_n\cdots i_1}_n=a^{i_n+}_n\cdots a^{i_1+}_1\psi_0 \qquad \mbox{and} \qquad \overline{\psi}^{i_n\cdots i_1}_n=\overline{a}^{i_n+}_n\cdots \overline{a}^{i_1+}_1\overline{\psi}_0.
\end{gather*}
These are the eigenfunctions of the Hamiltonians of the harmonic oscillator:
\begin{gather}
\label{eigwaarden}
h\psi^{i_n\cdots i_1}_n=\frac{\mu}{2}\frac{[n+m/2]_{q^2}}{q^{n+m/2}}\psi^{i_n\cdots i_1}_n \qquad \mbox{and} \qquad h^\ast\overline{\psi}^{i_n\cdots i_1}_n=\frac{\mu}{2}\frac{[n+m/2]_{q^2}}{q^{n+m/2}}\overline{\psi}^{i_n\cdots i_1}_n.
\end{gather}
Note that the functions $\psi^{i_n\cdots i_1}_n$ are not linearly independent.
These two types of functions generate vector spaces, which we denote by $\Pi(\mV)$ and $\overline{\Pi}(\mV)$. They are both representation of the abstract vector space $\mV$ which consists of linear combinations of abstract elements $\Psi_n^{i_n\cdots i_1}$. The maps $\Pi:\mV\to\Pi(\mV)$ and $\overline{\Pi}:\mV\to\overline{\Pi}(\mV)$ given by
\begin{gather*}
\Pi(\Psi^{i_n\cdots i_1}_n)=\psi^{i_n\cdots i_1}_n \qquad \mbox{and}\qquad \overline{\Pi}(\Psi^{i_n\cdots i_1}_n)=\overline{\psi}^{i_n\cdots i_1}_n
\end{gather*}
are isomorphisms. The vector space $\mV$ is an inner product space with the inner product $\langle\cdot|\cdot\rangle$ developed in \cite{MR1239953},
\begin{gather}
\label{inFiore}
\langle u,v\rangle = \int_{\gamma\cdot\mR^m_q}\left(\overline{\Pi}(u)\right)^\ast \Pi(v)+\left({\Pi}(u)\right)^\ast \overline{\Pi}(v).
\end{gather}
The value of~$\gamma$ is not important. The harmonic oscillator~$H$ on~$\mV$ is def\/ined such that $h\circ \Pi=\Pi\circ H$ and $h^\ast\circ \overline{\Pi}=\overline{\Pi}\circ H$ and is hermitian with respect to the inner product. The closure of~$\mV$ with respect to the topology induced by $\langle\cdot|\cdot\rangle$ is denoted by~$\cH$.

The behavior of the Fourier transforms with respect to the Hamiltonians of the harmonic oscillator was obtained in equation~\eqref{fourham}. This can be ref\/ined to the raising operators.
\begin{lemma}
\label{Foura}
For the Fourier transform $\cF^\pm_{\mR^m_q}$ in Definition~{\rm \ref{defFour}}, the expression $\cF^\pm_{\mR^m_q}\circ a^{j+}_n:\cV_{[\alpha]}\to\cV_{[q/\alpha]}$ satisfies
\begin{gather*}
\cF^\pm_{\mR^m_q}\circ a^{j+}_n = \pm iq^{2-n}\frac{b_n(q)}{\overline{b}_n(q)}\, \overline{a}^{j+}_n \circ \cF^\pm_{\mR^m_q}.
\end{gather*}
\end{lemma}

\begin{proof}
A direct calculation or the equations in the proof of Theorem~\ref{Fourinv} yield $\cF^\pm_{\mR^m_q}\circ\Lambda^{-1/4} =q^{m/2}\Lambda^{1/4}\circ \cF^\pm_{\mR^m_q}$. Combining this with Theorem~\ref{Fourierx} then yields the lemma.
\end{proof}

It is clear that $\Pi(\mV)\subset \cV_{[q^{-\frac{m}{2}}]}\oplus\cV_{[q^{1-\frac{m}{2}}]}$. It can be easily checked that the sum of $\cV_{[q^{-\frac{m}{2}}]}$ and $\cV_{[q^{1-\frac{m}{2}}]}$ is in fact direct. This implies that the Fourier transform can be trivially def\/ined on $\Pi(\mV)$. The Fourier transform of $\phi\in\Pi(\mV)$, with the unique decomposition $\phi=f+g$ with $f\in\cV_{[q^{-\frac{m}{2}}]}$ and $g\in\cV_{[q^{1-\frac{m}{2}}]}$, is def\/ined as
\begin{gather}
\label{defFourtss}
\cF^\pm_{\mR^m_q}(\phi) = \cF^\pm_{\mR^m_q}(f)+\cF^\pm_{\mR^m_q}(g)
\end{gather}
with the right hand side given in Def\/inition~\ref{defFour}. Now we def\/ine the Fourier transform on $\mV$.

\begin{definition}
The Fourier transform $\mF^\pm:\mV\to\mV$ is given by
\begin{gather*}
\mF^{\pm} = q^{-\frac{m^2}{4}}\overline{\Pi}^{-1}\circ\cF^\pm_{\mR^m_q}\circ\Pi,
\end{gather*}
with $\cF^\pm_{\mR^m_q}$ the Fourier transform on $\Pi(\mV)$ as in equation~\eqref{defFourtss}.
\end{definition}

Now we impose the condition $\overline{b}_n(q)=q^{2-n}b_n(q)$ on the undetermined coef\/f\/icients $b_n$, $\overline{b_n}$.

\begin{theorem}
\label{Foureig}
The Fourier transform $\mF^\pm$ on $\mV$ satisfies
\begin{gather*}
\mF^\pm[\Psi_n^{i_n\cdots i_1}] = (\pm i)^n\Psi_n^{i_n\cdots i_1}.
\end{gather*}
\end{theorem}

\begin{proof}
The def\/inition of $\mF^\pm$ shows that this statement is equivalent with $\cF^\pm_{\mR^m_q}[\psi^{i_n\cdots i_1}_n]=(\pm i)^n q^{\frac{m^2}{4}}\overline{\psi}^{i_n\cdots i_1}_n$. The proof of Theorem~\ref{Fourinv} implies $\cF^\pm_{\mR^m_q}[\psi_0]=q^{\frac{m^2}{4}}\overline{\psi}_0$. Lemma~\ref{Foura} then proves the theorem by induction.
\end{proof}

This immediately implies the following conclusions.
\begin{corollary}
\label{Pars}
The Fourier transform on $\mV$ can be continuously extended to $\cH$ and satisfies
\begin{gather*}
\mF^{\mp}\circ\mF^\pm = id_{\cH}
\end{gather*}
and the Parseval theorem
\begin{gather*}
\langle \mF^\pm(f)|\mF^{\pm}(g)\rangle = \langle f|g\rangle\qquad\mbox{for}\quad f,g\in\cH.
\end{gather*}
\end{corollary}

\begin{corollary}
The Fourier transform on $\cH$ can be written symbolically as
\begin{gather*}
\mF^\pm = \exp\left(\pm i\frac{\pi}{2}\left[\frac{{\rm arcsinh}\left(\frac{1-q^2}{\mu}H\right)}{\ln\big( \frac{1}{q}\big)}-\frac{m}{2}\right]\right).
\end{gather*}
\end{corollary}
\begin{proof}
This identity follows from evaluating the expression on $\Psi_n^{i_n\cdots i_1}$ using equation~\eqref{eigwaarden}.
\end{proof}

The Fourier transform $\cF^\pm_{\mR^m_q}$ on $\Pi(\mV)$ can therefore be written as
\begin{gather*}
\cF^\pm_{\mR^m_q} = q^{\frac{m^2}{4}}\overline{\Pi}\circ\Pi^{-1}\circ\exp\left(\pm i\frac{\pi}{2}\left[\frac{{\rm arcsinh}\left(\frac{1-q^2}{2\mu}\left[\bold{x}^2-\Delta\right]\right)}{\ln\big( \frac{1}{q}\big)}-\frac{m}{2}\right]\right).
\end{gather*}

The Parseval theorem in Corollary \ref{Pars} and the inner product~\eqref{inFiore} imply the following relation for $f,g\in\Pi(\mV)$:
\begin{gather*}
 q^{\frac{m^2}{4}}\int_{\gamma\cdot\mR^m_q}\left(\overline{\Pi}\circ\Pi^{-1}(f)\right)^\ast g+\left(f\right)^\ast \overline{\Pi}\circ\Pi^{-1}(g)\\
\qquad {} = \int_{\gamma\cdot\mR^m_q}\big(\cF^\pm_{\mR^m_q} (f)\big)^\ast \Pi\circ\overline{\Pi}^{-1}(\cF_{\mR^m_q}^\pm(g))+\big({\Pi}\circ\overline{\Pi}^{-1}(\cF^\pm_{\mR^m_q}(f))\big)^\ast \cF^\pm_{\mR^m_q}(g).
\end{gather*}
The results in \cite{MR1239953} show that this can be ref\/ined to
\begin{gather*}
q^{\frac{m^2}{4}}\int_{\gamma\cdot\mR^m_q}\left(\overline{\Pi}\circ\Pi^{-1}(f)\right)^\ast g = \int_{\gamma\cdot\mR^m_q}\big(\cF^\pm_{\mR^m_q} (f)\big)^\ast \Pi\circ\overline{\Pi}^{-1}(\cF_{\mR^m_q}^\pm(g)).
\end{gather*}

\section[The $q$-Fourier transform on Euclidean space]{The $\boldsymbol{q}$-Fourier transform on Euclidean space}
\label{qFeucl}

The techniques developed in this paper can also be applied to the theory of the $q$-Dirac operator on undeformed Euclidean space $\mR^m$, see~\cite{MR2595267}. We consider the polynomials in $m$ commuting variables: $\mC[x_1,\dots,x_m]$. The classical Laplace operator and norm squared are given by
\begin{gather*}
\Delta=\sum_{j=1}^m\partial^2_{x_j} \qquad \mbox{and} \qquad r^2=\sum_{j=1}^mx_j^2.
\end{gather*}
The spherical harmonics are the homogeneous null-solutions of the Laplace operator, $\cH_k=\mR[x_1,\dots,x_m]_k\cap {\rm Ker}\, \Delta$. The $q$-Fourier transforms of a function $H_k(\ux)\psi(r^2)$ with $H_k\in\cH_k$ are given by
\begin{gather*}
\overline{\cF}^{\pm}_{q}\big[H_k(\ux)\psi\big(r^2\big)\big](\underline{y}) = (\pm i)^kH_k(\uy)\int_0^{\frac{1}{\sqrt{1-q}}}d_qr\frac{J_{\frac{m}{2}+k-1}^{(1)}(rr_{\uy}|q^2)}{(rr_{\uy})^{\frac{m}{2}+k-1}}r^{m+2k-1}\psi\big(r^2\big),\\
\cF^{\pm}_{q}\big[H_k(\underline{y})\psi\big(r_{\uy}^2\big)\big](\ux) = (\pm i)^kH_k(\ux)\frac{1}{d\big(\frac{\gamma}{\sqrt{1+q}},\frac{m}{2}\big)}\int_0^{\gamma\cdot\infty}d_qr_{\uy}\frac{J_{\frac{m}{2}+k-1}^{(2)}
(qrr_{\uy}|q^2)}{(rr_{\uy})^{\frac{m}{2}+k-1}}r_{\uy}^{m+2k-1}\psi(r_{\uy}).
\end{gather*}
The fact that these transforms are each others inverse when evaluated on the appropriate function spaces follows immediately from equations \eqref{preHank1} and~\eqref{preHank2}. The operators~$D_j$, $j=1,\dots,m$ are def\/ined in~\cite[equation~(25) and Def\/inition~2]{MR2595267}. They are a $q$-deformation of the partial derivatives on~$\mR^m$. Similarly to equation~\eqref{partialr} for the partial derivatives on $q$-Euclidean space the $q$-derivatives on Euclidean space satisfy{\samepage
\begin{gather*}
D_if(r^2)=(1+q)x_i\big(\partial_{r^2}^{q^2}f\big(r^2\big)\big)+f\big(q^2r^2\big)D_i,
\end{gather*}
see Lemma~6 in \cite{MR2595267}.}

The $q$-Laplace operator on $\mR^m$ can be def\/ined as $ \Delta_q=\sum\limits_{j=1}^mD_j^2$. The polynomial null-solution of this $q$-Laplace operator correspond to the classical spaces $\cH_k$. This shows that the $q$-deformation is purely radial. This can also be seen from the fact that the Howe dual pair of this construction is $(O(m),\cU_q(\mathfrak{sl}_2))$. This implies that there is no spherical deformation and a~radial deformation identical to the one from quantum Euclidean space.

Using the results from the previous sections and~\cite{MR2595267} it is straightforward to prove that
\begin{gather*}
D_j\overline{\cF}^{\pm }_q[f(\ux)] = \pm i\overline{\cF}^{\pm}_q[x_jf(\ux)]
\end{gather*}
holds. Also the theory of the harmonic oscillator and the corresponding Hilbert space can be trivially translated to this setting.

\subsection*{Acknowledgements}
The author would like to thank Hendrik De Bie for helpful suggestions and comments.

\pdfbookmark[1]{References}{ref}
\LastPageEnding

\end{document}